\documentclass[a4paper,11pt]{article}
\pdfoutput=1 
\usepackage{jheppub,bm} 
\usepackage[T1]{fontenc} 
\usepackage{slashed}
\usepackage{soul}
\usepackage{amsmath}
\newcommand{\bef}{\begin{figure}[hbt]\centering}
\newcommand{\eef}{\end{figure}}
\allowdisplaybreaks

\usepackage{graphicx}
\def\BNU{Center of Advanced Quantum Studies, Department of Physics, Beijing Normal University, Beijing 100875, China}
\def\LANL{Theoretical Division, Los Alamos National Laboratory, Los Alamos, New Mexico 87545, USA}
\def\ucla{Department of Physics and Astronomy, University of California, Los Angeles, California 90095, USA}
\def\theory{Mani L. Bhaumik Institute for Theoretical Physics, University of California, Los Angeles, California 90095, USA}
\def\Northwest{Department of Physics and Astronomy, Northwestern University, Evanston, Illinois 60208, USA} 
\def\ANL{High Energy Physics Division, Argonne National Laboratory, Argonne, Illinois 60439, USA}
\def\LBL{Nuclear Science Division, Lawrence Berkeley National Laboratory, Berkeley, California 94720, USA}

\title{The transverse momentum distribution of hadrons within jets}

\author{Zhong-Bo Kang$^{a,b,c}$, }    
\author{Xiaohui Liu$^{d}$, }
\author{Felix Ringer$^{e}$}
\author{and Hongxi Xing$^{f,g}$}
 
\affiliation[a]{\ucla}
\affiliation[b]{\theory}
\affiliation[c]{\LANL}
\affiliation[d]{\BNU}
\affiliation[e]{\LBL}
\affiliation[f]{\Northwest}
\affiliation[g]{\ANL}

\emailAdd{zkang@physics.ucla.edu}
\emailAdd{xiliu@bnu.edu.cn}
\emailAdd{fmringer@lbl.gov}
\emailAdd{hxing@northwestern.edu}
\newcommand{\GG}{{\cal G}}
\newcommand{\nnu}{\nonumber\\}

\def\C{{\tilde C}}

\newcommand{\kt}{anti-k$_{\rm T}$}

\newcommand{\beq}{\begin{equation}}
\newcommand{\eeq}{\end{equation}}
\def\bea#1\eea{\begin{align}#1\end{align}}
\newcommand{\be}{\begin{eqnarray}}
\newcommand{\ee}{\end{eqnarray}}

\newcommand{\h}{{\mathcal H}}

\newcommand{\sla}[1]{{#1}\!\!\!\slash}

\abstract{We study the transverse momentum distribution of hadrons within jets, where the transverse momentum is defined with respect to the standard jet axis. We consider the case where the jet substructure measurement is performed for an inclusive jet sample $pp\to\text{jet}+X$. We demonstrate that this observable provides new opportunities to study transverse momentum dependent fragmentation functions (TMDFFs) which are currently poorly constrained from data, especially for gluons. The factorization of the cross section is obtained within Soft Collinear Effective Theory (SCET), and we show that the relevant TMDFFs are the same as for the more traditional processes semi-inclusive deep inelastic scattering (SIDIS) and electron-positron annihilation. Different than in SIDIS, the observable for the in-jet fragmentation does not depend on TMD parton distribution functions which allows for a cleaner and more direct probe of TMDFFs. We present numerical results and compare to available data from the LHC.}

\begin{document} 
\maketitle
\flushbottom

\section{Introduction}
In recent years, studies of jets and their internal structure have played increasingly important roles in testing the fundamental properties of Quantum Chromodynamics (QCD), and in searching for new physics beyond the Standard Model~\cite{Altheimer:2013yza,Adams:2015hiv}. 
This is the case in particular in the era of the Large Hadron Collider (LHC), where collimated jets of hadrons are abundantly produced. 

In this paper we study the transverse momentum distribution of hadrons $h$ within fully reconstructed jets in $pp$ collisions, $pp\to (\mathrm{jet}h)X$, as illustrated in Fig.~\ref{fig:JFF}. Specifically we study the ratio
\bea
F(z_h, {\bm j}_\perp; \eta, p_T, R) = \left.\frac{d\sigma^{pp\to (\text{jet}\,h)X}}{dp_Td\eta dz_h d^2 {\bm j}_\perp}
\right/\frac{d\sigma^{pp\to\text{jet}X}}{dp_Td\eta }\,,
\eea
where the numerator and denominator are the differential jet cross sections with and without the reconstruction of the hadron $h$ inside the jet. The variables $\eta$, $p_T$ and $R$ are the rapidity, the transverse momentum and the jet size parameter of the reconstructed jet measured in the center-of-mass (CM) frame in $pp$ collisions. The large light-cone momentum fraction of the jet carried by the hadron $h$ is denoted by $z_h$ and ${\bm j}_\perp$ is the transverse momentum of the hadron with respect to the standard jet axis. Throughout this paper, bold letters represent two-dimensional transverse momentum vectors, whereas the magnitude of these vectors is referred to as, for example, $j_\perp = |{\bm j}_\perp|$. This observable has been measured at the LHC in $pp$ collisions for a wide range of jet transverse momenta $p_T$~\cite{Aad:2011sc}. In addition, it has been measured in both unpolarized $pp$ and transversely polarized $p^{\uparrow}p$ collisions at the Relativistic Heavy Ion Collider (RHIC)~\cite{Aschenauer:2013woa,Aschenauer:2015eha,Aschenauer:2016our}. It was proposed in \cite{Yuan:2007nd} that the latter case can be used to probe azimuthal spin correlations in the fragmentation process, in particular, the so-called Collins function~\cite{Collins:1992kk}. 
\bef
\includegraphics[width=1.1in]{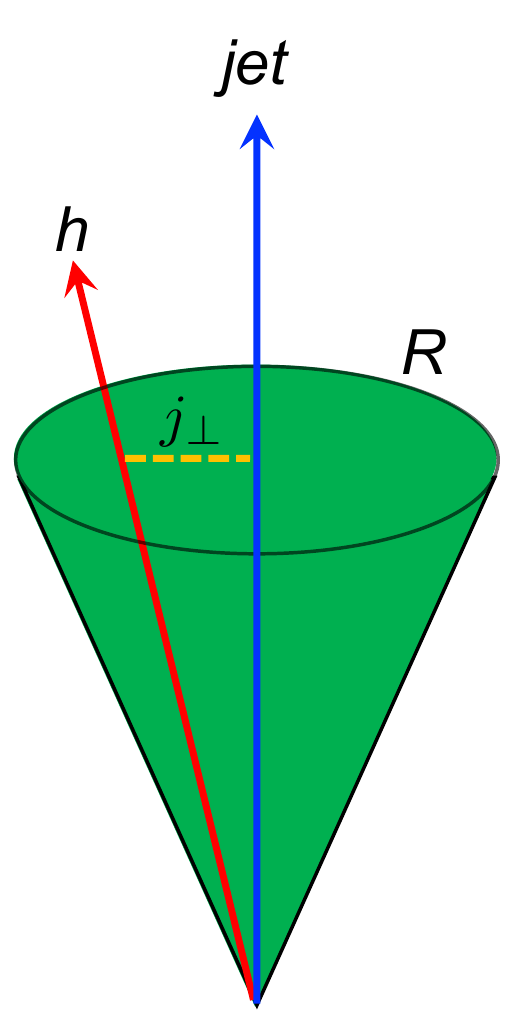}
\caption{The Distribution of hadrons inside a fully reconstructed jet. Here, ${\bm j}_\perp$ is the transverse momentum of hadrons with respect to the standard jet axis, and $R$ is the jet radius.}
\label{fig:JFF} 
\eef

In this work, we develop the theoretical framework to study the above observable $F(z_h, {\bm j}_\perp; \eta, p_T, R)$. We consider the case where the jet substructure measurement is performed for an inclusive jet sample $pp\to\text{jet}+X$, different than the study in \cite{Bain:2016rrv} where an exclusive jet sample was studied in the context of heavy quarkonium production. As the experimental measurements~\cite{Aad:2011sc} were performed for inclusive jet samples, our approach facilitates a direct comparison with the experimental data. In particular, we concentrate on the region of the hadron transverse momentum where $j_\perp \ll p_T R$. Here, $j_\perp$ is defined with respect to the standard jet axis, rather than a recoil-free axis, specifically the winner-take-all jet axis as discussed in~\cite{Neill:2016vbi}. While a recoil-free axis can be advantageous for various applications for collider physics, it turns out that there is only a direct relation to the standard transverse momentum dependent fragmentation functions (TMDFFs) when the standard jet axis is used. The standard TMDFFs are also probed in the traditional processes semi-inclusive deep inelastic scattering (SIDIS) and back-to-back hadron pair production in electron-positron annihilation.

Following earlier work on the longitudinal momentum distribution of hadrons inside jets~\cite{Procura:2009vm,Jain:2011xz,Arleo:2013tya,Kaufmann:2015hma,Chien:2015ctp,Kang:2016ehg}, we can write down the factorized form of the cross section in $pp$ collisions as follows (for more details, see Eq.~\eqref{eq:sigjethX} below)
\bea
\label{eq:master}
\frac{d\sigma^{pp\to (\text{jet}\,h)X}}{dp_Td\eta dz_h d^2 {\bm j}_\perp}=\sum_{a,b,c}f_a(x_a,\mu)\otimes f_b(x_b,\mu)\otimes H_{ab}^c(x_a,x_b,\eta,p_T/z,\mu)\otimes {\cal G}_c^h(z,z_h,\omega_J R,{\bm j}_\perp,\mu)\,.
\eea
Here, $f_{a,b}$ denote the parton distribution functions (PDFs) in the proton with the corresponding momentum fraction $x_a$ and $x_b$, respectively. The hard functions $H^c_{ab}$ describe the production of an energetic parton $c$ in the hard-scattering event. In addition, the functions ${\cal G}_c^h(z,z_h,\omega_J R,{\bm j}_\perp,\mu)$ are the semi-inclusive TMD fragmenting jet functions (siTMDFJFs) which we describes the production of a jet in the final state with the observed hadron inside. We define this new function in Sec.~\ref{sec:fac} below. We further demonstrate that the siTMDFJFs can be refactorized in terms of hard matching functions, soft functions, and the transverse momentum dependent fragmentation functions. It is evident that the in-jet fragmentation of hadrons considered in this work provides a very sensitive probe especially for the gluon TMDFF which is so far only poorly constrained by the traditional processes.

The remainder of this paper is organized as follows. In Sec.~\ref{sec:fac}, we provide operator definitions for the siTMDFJFs ${\cal G}_c^h$, which is the essential component in describing the hadron transverse momentum distribution inside jets. We derive a factorization formalism for siTMDFJFs in terms of hard functions, soft functions and the TMDFFs. We calculate all these relevant functions in the factorized expression to next-to-leading order (NLO) and derive their renormalization group equations. We solve the resulting renormalization group (RG) equations in order to resum all the relevant large logarithms. In Sec.~\ref{sec:pheno}, we provide a first numerical estimate for the hadron transverse momentum distribution inside jets for LHC kinematics, and we compare with experimental results. We summarize our paper and provide further discussions in Sec.~\ref{sec:sum}.

\section{The semi-inclusive TMD fragmenting jet function}\label{sec:fac}
In this section, we introduce the definition of the semi-inclusive TMD fragmenting jet functions $\GG_{c}^h(z, z_h, \omega_J R, {\bm j}_\perp, \mu)$, which are the relevant new ingredients in order to describe the hadron transverse momentum distribution within jets produced in $pp$ collisions. The siTMDFJFs describe the fragmentation of a hadron $h$ inside a jet that is initiated by a parton $c$. We first provide their operator definitions, and we then derive the factorization formalism in terms of hard functions, soft functions and TMDFFs. We derive the relevant RG equations and their solutions. In addition, we work out the relation to standard TMDFFs probed in SIDIS and electron-positron annihilation.

\subsection{Definition}
Following the convention for the definition of the semi-inclusive fragmenting jet function in~\cite{Kang:2016ehg}, the siTMDFJFs are defined for quark and gluon jets as follows
\begin{subequations}
\label{eq:def}
\bea
\GG_q^h(z, z_h, \omega_J R, {\bm j}_\perp, \mu) =& \frac{z}{2N_c} \delta\left(z_h - \frac{\omega_h}{\omega_J}\right)
{\rm Tr}
\bigg[\frac{\sla{\bar n}}{2}
\langle 0| \delta\left(\omega - \bar n\cdot {\mathcal P} \right)  \delta^2\left({\mathcal P}_\perp - {\bm j}_\perp \right) \chi_n(0)  |(Jh)X\rangle
\nnu
&\hspace{30mm}  \times
\langle (Jh)X|\bar \chi_n(0) |0\rangle \bigg],
\\
\GG_g^h(z, z_h, \omega_J R, {\bm j}_\perp, \mu) =& - \frac{z\,\omega}{(d-2)(N_c^2-1)} \delta\left(z_h - \frac{\omega_h}{\omega_J}\right) \langle 0|  \delta\left(\omega - \bar n\cdot {\mathcal P} \right) \delta^2\left({\mathcal P}_\perp - {\bm j}_\perp \right)
\nnu
&\hspace{30mm} \times
   {\mathcal B}_{n\perp \mu}(0) 
 |(Jh)X\rangle \langle (Jh)X|{\mathcal B}_{n\perp}^\mu(0)  |0\rangle,
\eea
\end{subequations}
where $N_c$ is the number of the colors for quarks, and ${\bm j}_\perp$ is the transverse momentum of the hadron $h$ with respect to the standard jet axis. The large light-cone momentum components of the initial parton $c$, the jet, and the hadron $h$ are given by $\omega, ~\omega_J, ~\omega_h$, respectively. We choose to express the results of our calculation in terms of the following ratios of these variables 
\bea
z = \frac{\omega_J}{\omega},
\qquad
z_h = \frac{\omega_h}{\omega_J},
\eea
as well as $\omega_J$ which is related to the transverse momentum of the reconstructed jet. In Eq.~\eqref{eq:def}, we have $n^\mu=(1, \hat n)$ and $\bar n^\mu = (1, -\hat n)$ where the spatial component $\hat n$ is chosen along the standard jet axis. In addition, we have $n^2=\bar n^2=0$ and $n\cdot \bar n=2$. The gauge invariant collinear quark and gluon fields within 
Soft Collinear Effective Theory (SCET)~\cite{Bauer:2000ew,Bauer:2000yr,Bauer:2001ct,Bauer:2001yt,Bauer:2002nz}  are denoted by $\chi_n$ and ${\mathcal B}_{n\perp}^\mu$ as usual, and ${\mathcal P}$ is the label momentum
operator. The sum over states $|X\rangle$ runs over all final state particles except for the observed jet $J$ with the identified hadron $h$ inside.

In~\cite{Kang:2016mcy,Kang:2016ehg}, it was found that the characteristic momentum scale for the jet dynamics with a jet radius $R$ is given by
\bea
\mu_J \sim \omega_J\tan(R/2) \,.
\label{eq:jetscale}
\eea
We would like to point out that for the standard jet algorithms in $pp$ collisions, one can simply make the replacement $\omega_J\tan(R/2)\to p_TR$, where the jet size $R$ is defined in the $(\eta, \phi)$ plane, see e.g. Ref.~\cite{Mukherjee:2012uz,Kang:2017mda}. Depending on the relative scaling of $j_\perp$, $\mu_J$ and $\Lambda_{\text{QCD}}$ one finds different factorization theorems for the hadron-in-jet cross section. 

First, we consider the case when $j_\perp$ is of the same order as the characteristic jet scale $\mu_J$, i.e., $\Lambda_{\rm QCD} \ll j_\perp\sim p_T R $. In this case, the siTMDFJF $\GG_{c}^h$  as defined in Eq.~\eqref{eq:def} can be factorized into standard collinear fragmentation functions $D_{h/i}(z_h, \mu)$  convolved with Wilson coefficients in $z_h$~\cite{Jain:2011iu,Bain:2016rrv}. The Wilson coefficients are functions of $z$, $\omega_J$ and $j_\perp$, and can be calculated perturbatively.

Second, in the region where $j_\perp$ is much smaller than the characteristic jet scale $\mu_J$, i.e., $\Lambda_{\rm QCD} \lesssim j_\perp \ll p_T R $, the perturbative expansion is plagued with large logarithmic corrections of the form 
$\ln\left( p_TR/j_\perp\right)$, and the standard collinear factorization breaks down. Therefore, in the small $j_\perp$ regime, a new factorization formalism -- TMD factorization~\cite{Collins:2011zzd}, is required to recover reliable predictions within QCD perturbation theory. This is the kinematic region that we address in this work.

\subsection{Factorization theorem}
We focus on the kinematic region with the relative scalings $\Lambda_{\rm QCD}\lesssim j_\perp \ll p_TR$, referred to as a TMD region. In this region, since the transverse momentum $j_\perp$ inside the jet is parametrically small, only collinear radiation within the jet with the momentum scaling $p_{c}=(p_c^-, p_c^+, p_{c\perp})\sim p_T  (1, \lambda^2, \lambda)$ where $\lambda\sim j_\perp/p_T$, and the soft radiation of order $j_\perp $ are relevant to leading power of the cross section.~\footnote{We note that the soft degrees of freedom considered here are the same as the coft or c-soft modes introduced in~\cite{Becher:2015hka} and~\cite{Chien:2015cka}, respectively.}
Harder emissions are only allowed outside the jet and, therefore, do not affect the hadron transverse momentum $j_\perp$.  
Since $j_\perp$ is defined with respect to the jet axis, any radiation outside the jet will only influence the determination of the jet axis but have no impact on the $j_\perp$ spectrum. Therefore, we have the following factorized form for $\GG_{c}^h$ derived within SCET
\bea
\GG_{c}^h(z, z_h, \omega_J R, {\bm j}_\perp, \mu)  =& \h_{c\to i}(z, \omega_J R, \mu) 
\int d^2 {\bm k}_\perp d^2{\bm \lambda}_\perp \delta^2\left(z_h {\bm \lambda}_\perp + {\bm k}_\perp - {\bm j}_\perp\right)
\nnu
&\times D_{h/i}(z_h, {\bm k}_\perp, \mu, \nu) S_i({\bm \lambda}_\perp, \mu, \nu R)\; ,
\label{eq:GG}
\eea
where besides the usual renormalization scale $\mu$, the scale $\nu$ arises due to the so-called rapidity divergences in the relevant functions to be discussed in detail below. Here, $\h_{c\to i}(z, \omega_J R, \mu)$ are hard matching functions related to out-of-jet radiation. The soft functions $S_i({\bm \lambda}_\perp, \mu, \nu R)$ take into account soft radiation inside the jet and $D_{h/i}(z_h, {\bm k}_\perp, \mu, \nu)$ are the usual TMDFFs, which characterize the collinear degrees of freedom inside the jet. 

The delta function relates the observed transverse momentum of the hadron ${\bm j}_\perp$ to be the total transverse momentum of soft and collinear radiations. Note that ${\bm \lambda}_\perp$ is multiplied by $z_h$ inside the delta function to account for the difference between the fragmenting parton and the observed hadron with respect to the jet axis. All the ingredients in the factorization formula will be calculated up to NLO, which determines their RG evolutions. All large logarithms of the form $ \ln R$ and $\ln \left(p_T R/j_\perp\right)$ are resummed by solving the obtained RG equations and by running each component from their natural scales to the hard scale $\mu \sim p_T$ at which the cross section is evaluated. 

\subsection{Hard functions}
The hard matching functions $\h_{c\to i}(z, \omega_J R, \mu)$ encode radiation with a virtuality of order ${\cal O}(p_T R)$ outside of the jet.
They describe how an energetic parton $c$ from the hard-scattering event produces a jet initiated by parton $i$ with energy $\omega_J$ and radius $R$, and can be computed through the matching relation in Eq.~\eqref{eq:GG}. Up to NLO, they are obtained by the out-of-jet radiation diagrams for inclusive jet (substructure) observables~\cite{Kang:2016ehg}. The same hard matching functions were found in the context of central subjets measured on an inclusive jet sample in~\cite{Kang:2017mda}. For \kt ~jets, the renormalized expressions are given by
\begin{subequations}
\label{eq:hard}
\bea
\h_{q\to q'}(z,\omega_J R,\mu) 
  &= \delta_{qq'}\delta(1-z) +\delta_{qq'} \frac{\alpha_s}{2\pi} \bigg[C_F \delta(1-z)\Big(-\frac{L^2}{2} - \frac{3}{2} L +\frac{\pi^2}{12} \Big) 
\nnu
 & \quad
+ P_{qq}(z) L -2C_F(1+z^2)\left(\frac{\ln(1-z)}{1-z}\right)_+ -C_F(1-z)  \bigg] \,, 
\\[.2cm]
 \h_{q\to g}(z,\omega_J R,\mu) 
 &=\frac{\alpha_s}{2\pi}\bigg[\Big(L - 2 \ln(1-z) \Big) P_{gq}(z) - C_Fz \bigg]\,, 
 \\[.2cm]
\h_{g\to g}(z, \omega_J R, \mu) 
& = \delta(1-z) + \frac{\alpha_s}{2\pi}\bigg[ \delta(1-z)\Big(-C_A\frac{L^2}{2} - \frac{\beta_0}{2} L + \frac{\pi^2}{12}\Big)
\nnu
& \quad
+ P_{gg}(z) L - \frac{4C_A (1-z+z^2)^2}{z} \left(\frac{\ln(1-z)}{1-z}\right)_{+} \bigg]\,,
\\[.2cm]
\h_{g\to q}(z,\omega_J R, \mu) 
 & =  \frac{\alpha_s}{2\pi}\bigg[\Big(L - 2\ln(1-z) \Big)  P_{qg}(z) - T_F 2z(1-z) \bigg]\,, 
\eea
\end{subequations}
where the logarithm $L$ is defined as
\bea
L = \ln\left( \frac{ \mu^2} { \omega_J^2 \tan^2 (R/2)} \right),
\eea
and the standard splitting functions are also provided here for reference
\bea
P_{qq}(z) &= C_F \left[\frac{1+z^2}{(1-z)_+} + \frac{3}{2}\delta(1-z) \right]\,,
\label{eq:Pqq}
\\[.2cm]
P_{gq}(z) &= C_F \frac{1+(1-z)^2}{z}\,,
\\[.2cm]
P_{gg}(z) &= 2C_A \left[\frac{z}{(1-z)_+} + \frac{1-z}{z} +z(1-z) \right] + \frac{\beta_0}{2} \delta(1-z)\,,
\label{eq:Pgg}
\\[.2cm]
P_{qg}(z) &= T_F\left[z^2+(1-z)^2\right]\,.
\eea
The analogous hard matching coefficients for cone jets can be found in~\cite{Kang:2017mda}. The RG equations of the functions $\h_{i\to j}$ take the following form
\bea
\label{eq:RG_H}
\mu\frac{d}{d\mu} \h_{i\to j}(z, \omega_J R, \mu) =  \sum_k \int_z^1 \frac{dz'}{z'} \gamma_{ik}\left(\frac{z}{z'},\omega_J R,\mu\right) \h_{k\to j}(z', \omega_J R, \mu)\,.
\eea
Note that this is a set of four coupled equations with a DGLAP type structure. The anomalous dimensions $\gamma_{ij}(z,\omega_J R,\mu)$ are given by
\bea
\label{eq:gamma_H}
\gamma_{ij}(z,\omega_J R,\mu) = \delta_{ij} \delta(1-z) \Gamma_i(\omega_J R,\mu) + \frac{\alpha_s}{\pi} P_{ji}(z)\,.
\eea
The second terms of the $\gamma_{ij}(z,\omega_J R,\mu)$ are the standard DGLAP evolution kernels. Instead, the first term contains a logarithm $L$ and the functions $\Gamma_i(\omega_J R,\mu)$ are given at this order by
\begin{subequations}
\label{eq:Gamma_i}
\bea
\Gamma_q(\omega_J R,\mu) &= \frac{\alpha_s}{\pi} C_F\left(- L - \frac{3}{2}\right),
\\
\Gamma_g(\omega_J R,\mu) &= \frac{\alpha_s}{\pi} C_A  \left( - L - \frac{\beta_0}{2C_A}\right).
\eea
\end{subequations}
To summarize, the RG equations encountered here resum double logarithms whereas the DGLAP equations are always associated with the resummation of single logarithms. Evidently, the natural scale for the hard matching coefficients $\h_{i\to j}(z, \omega_J R, \mu)$ is the same as the jet scale $\mu_J$ as defined in Eq.~\eqref{eq:jetscale}, i.e.,
\bea
\mu_J \sim \omega_J  \tan (R/2) \to p_T R \,.
\eea
Thus, by solving the above RG equations and by evolving the hard matching functions from the scale $\mu_J \sim p_T R$ to the hard-scattering scale $\mu\sim p_T$ where the cross section is evaluated, we are resumming large logarithms of jet radius $\ln R$. 

\subsection{TMD fragmentation functions}
Now we focus on the TMDFFs $D_{h/i}(z_h, {\bm k}_\perp, \mu, \nu)$, which are defined as
\begin{subequations}
\label{eq:defD}
\bea
D_{h/q}(z_h, {\bm k}_\perp, \mu,\nu) =& \frac{z_h}{2N_c}
{\rm Tr} 
\bigg[\frac{\sla{\bar n}}{2}
\langle 0| \delta\left(\omega - \bar n\cdot {\mathcal P} \right) \delta^2\left({\mathcal P}_\perp - {\bm k}_\perp \right)  \chi_n(0)  |(Jh)X\rangle
\nnu
&\hspace{10mm}  \times
\langle (Jh)X|\bar \chi_n(0) |0\rangle \bigg],
\\
D_{h/g}(z_h, {\bm k}_\perp, \mu,\nu) =& - \frac{z_h\,\omega}{(d-2)(N_c^2-1)}
\langle 0|  \delta\left(\omega - \bar n\cdot {\mathcal P} \right) \delta^2\left({\mathcal P}_\perp - {\bm k}_\perp \right) {\mathcal B}_{n\perp \mu}(0) 
\nnu
&\hspace{10mm} \times
 |(Jh)X\rangle \langle (Jh)X|{\mathcal B}_{n\perp}^\mu(0)  |0\rangle,
\eea
\end{subequations}
where $\omega$ is the light-cone energy of the initiating quark or gluon. The TMDFFs contain rapidity divergences. We choose to employ the analytic rapidity regulator of~\cite{Chiu:2012ir} which introduces a dependence on the associated rapidity scale $\nu$. Traditionally, TMDs are studied conveniently in Fourier transform space or $b$-space. Following the standard convention of Ref.~\cite{Collins:2011zzd,Echevarria:2014xaa}, we define the TMDFFs in $b$-space as 
\bea
D_{h/i}(z_h, {\bm b}, \mu, \nu) = \frac{1}{z_h^2} \int d^2{\bm k}_\perp e^{-i{\bm k}_\perp\cdot{\bm b}/z_h} D_{h/i}(z_h, {\bm k}_\perp, \mu, \nu)\,. 
\label{eq:Fourier}
\eea
At the same time, through the inverse Fourier transform, we obtain the TMDFFs in momentum space as
\bea
\label{eq:inverse-Fourier}
D_{h/i}(z_h, {\bm k}_\perp, \mu, \nu) = \int \frac{d^2{\bm b}}{(2\pi)^2} e^{i{\bm k}_\perp\cdot {\bm b}/z_h} D_{h/i}(z_h, {\bm b}, \mu, \nu)\,.
\eea

\subsubsection{Perturbative results}
In perturbation theory, the bare TMDFFs suffer from infrared (IR), ultra-violet (UV), and rapidity divergences. To understand the features of these divergences, it is instructive to study the perturbative results for the TMDFFs in the region where $k_\perp\gg \Lambda_{\rm QCD}$. In the following, we consider perturbative splittings $i\to jk$ at the parton level, where $j$ refers to the identified parton whose transverse momentum ${\bm k}_\perp$ is measured. We denote the corresponding TMDFFs at the parton level by $D_{j/i}(z_h, {\bm k}_\perp, \mu, \nu)$. Up to NLO, we find the following results
\begin{subequations}
\bea
D_{q/q}(z_h, {\bm k}_\perp, \mu, \nu) =& \delta(1-z_h) \delta^2({\bm k}_\perp)+ \frac{\alpha_s}{2\pi^2} C_F \Gamma(1+\epsilon) e^{\epsilon\gamma_E} \frac{1}{\mu^2} \left(\frac{\mu^2}{ k_\perp^2}\right)^{1+\epsilon}
\nnu
&\times 
\left[\frac{2z_h}{(1-z_h)^{1+\eta}} \left(\frac{\nu}{\omega_J}\right)^\eta+(1-\epsilon) (1-z_h)\right],
\\[.3cm]
D_{g/q}(z_h, {\bm k}_\perp, \mu, \nu) = &  
\frac{\alpha_s}{2\pi^2} C_F \Gamma(1+\epsilon) e^{\epsilon\gamma_E} \frac{1}{\mu^2} \left(\frac{\mu^2}{k_\perp^2}\right)^{1+\epsilon} 
 \left[   \frac{1+(1-z_h)^2}{z_h}  - \epsilon \, z_h    \right],
\\[.3cm]
D_{g/g}(z_h, {\bm k}_\perp, \mu, \nu) =& \delta(1-z_h) \delta^2({\bm k}_\perp)+ \frac{\alpha_s}{2\pi^2} C_A \Gamma(1+\epsilon) e^{\epsilon\gamma_E} \frac{1}{\mu^2} \left(\frac{\mu^2}{k_\perp^2}\right)^{1+\epsilon}
\nnu
&\times 
z_h \left[\frac{1+z_h}{(1-z_h)^{1+\eta}} \left(\frac{\nu}{\omega_J}\right)^\eta+\left(3-2z_h\right)
+2 (1-\epsilon^2)\frac{1-z_h}{z_h^2}\right],
\\[.3cm]
D_{q/g}(z_h, {\bm k}_\perp, \mu, \nu) = &
 \frac{\alpha_s}{2\pi^2} T_F \Gamma(1+\epsilon) e^{\epsilon\gamma_E} \frac{1}{\mu^2} \left(\frac{\mu^2}{k_\perp^2}\right)^{1+\epsilon} 
 \left[1 - \frac{2z_h (1-z_h) }{1-\epsilon}\right]
\,. 
\eea
\end{subequations}
Note that the results for the TMDFFs here are independent of the jet radius parameter $R$. The allowed collinear radiation inside the jet is so collimated along the jet axis in the kinematical limit that we are considering, $j_\perp /{p_T} \ll R$, such that it is insensitive to the jet boundary. We now calculate the Fourier transform of these results according to Eq.~\eqref{eq:Fourier}. After expanding around $\eta\to 0$ and then $\epsilon\to 0$, we obtain the following results in $b$-space
\begin{subequations}
\label{eq:TMDb}
\bea
\label{eq:qq}
D_{q/q}(z_h, {\bm b}, \mu, \nu) =& \frac{1}{z_h^2}\Bigg\{\delta(1-z_h) 
\nnu
&
+ \frac{\alpha_s}{2\pi} C_F \bigg[
\frac{2}{\eta} \left(\frac{1}{\epsilon}+\ln\left(\frac{\mu^2}{\mu_b^2}\right)\right) +\frac{1}{\epsilon}\left(2\ln\left(\frac{\nu}{\omega_J}\right)+\frac{3}{2}\right)\bigg]\delta(1-z_h)
\nnu
&+\frac{\alpha_s}{2\pi} \bigg[-\frac{1}{\epsilon}-\ln\left(\frac{\mu^2}{z_h^2\mu_b^2}\right)\bigg]P_{qq}(z_h)
\nnu
&+\frac{\alpha_s}{2\pi} C_F \bigg[\ln\left(\frac{\mu^2}{\mu_b^2}\right) \left(2 \ln\left(\frac{\nu}{\omega_J}\right) + \frac{3}{2}\right) \delta(1-z_h) + (1-z_h)\bigg]\Bigg\},
\\[.2cm]
\label{eq:gq}
D_{g/q}(z_h, {\bm b}, \mu, \nu) =& \frac{1}{z_h^2}\Bigg\{
\frac{\alpha_s}{2\pi} \bigg[-\frac{1}{\epsilon}-\ln\left(\frac{\mu^2}{z_h^2\mu_b^2}\right)\bigg]P_{gq}(z_h)
+\frac{\alpha_s}{2\pi} C_F z_h\Bigg\},
\\[.2cm]
\label{eq:gg}
D_{g/g}(z_h, {\bm b}, \mu, \nu) =& \frac{1}{z_h^2}\Bigg\{\delta(1-z_h) 
\nnu
&
+ \frac{\alpha_s}{2\pi} C_A \bigg[
\frac{2}{\eta} \left(\frac{1}{\epsilon}+\ln\left(\frac{\mu^2}{\mu_b^2}\right)\right) +\frac{1}{\epsilon}\left(2\ln\left(\frac{\nu}{\omega_J}\right)+\frac{\beta_0}{2C_A}\right)\bigg]\delta(1-z_h)
\nnu
&+\frac{\alpha_s}{2\pi} \bigg[-\frac{1}{\epsilon}-\ln\left(\frac{\mu^2}{z_h^2\mu_b^2}\right)\bigg]P_{gg}(z_h)
\nnu
&+\frac{\alpha_s}{2\pi} C_A \bigg[\ln\left(\frac{\mu^2}{\mu_b^2}\right) \left(2 \ln\left(\frac{\nu}{\omega_J}\right) + \frac{\beta_0}{2C_A}\right) \delta(1-z_h)\bigg]\Bigg\},
\\[.2cm]
\label{eq:qg}
D_{q/g}(z_h, {\bm b}, \mu, \nu) =& \frac{1}{z_h^2}\Bigg\{
\frac{\alpha_s}{2\pi} \bigg[-\frac{1}{\epsilon}-\ln\left(\frac{\mu^2}{z_h^2\mu_b^2}\right)\bigg]P_{qg}(z_h)
+\frac{\alpha_s}{2\pi} T_F 2 z_h(1-z_h)\Bigg\}\,.
\eea
\end{subequations}
Here we introduced the scale $\mu_b$ which is defined as $\mu_b = 2e^{-\gamma_E}/b$~\cite{Aybat:2011zv}. 

\subsubsection{Renormalization}
In this section we perform the renormalization of the TMDFFs and derive the resulting RG equations. We observe that the poles of $D_{q/q}$ and $D_{g/g}$ in the second lines of Eqs.~\eqref{eq:qq} and \eqref{eq:gg} are UV poles. Therefore, they are subtracted via the usual renormalization procedure. The bare and renormalized TMDFFs are related as
\bea
D_{h/i}(z_h, {\bm b}, \mu, \nu) &= Z^D_i({\bm  b}, \mu, \nu) D_{h/i}^{\rm bare}(z_h, {\bm b}, \mu, \nu)\,.
\eea
For the relevant renormalization constants, we find
\begin{subequations}
\bea
Z^D_q({\bm  b}, \mu, \nu) &= 1+ \frac{\alpha_s}{2\pi} C_F \bigg[
\frac{2}{\eta} \left(\frac{1}{\epsilon}+\ln\left(\frac{\mu^2}{\mu_b^2}\right)\right) +\frac{1}{\epsilon}\left(2\ln\left(\frac{\nu}{\omega_J}\right)+\frac{3}{2}\right)\bigg]\,,
\\[.2cm]
Z^D_g({\bm  b}, \mu, \nu) &= 1+ \frac{\alpha_s}{2\pi} C_A \bigg[
\frac{2}{\eta} \left(\frac{1}{\epsilon}+\ln\left(\frac{\mu^2}{\mu_b^2}\right)\right) +\frac{1}{\epsilon}\left(2\ln\left(\frac{\nu}{\omega_J}\right)+\frac{\beta_0}{2C_A}\right)\bigg]\,.
\eea
\end{subequations}
We thus obtain the associated RG equation and the rapidity renormalization group (RRG) equation
\bea
\mu\frac{d}{d\mu} \ln D_{h/i}(z_h, {\bm b}, \mu, \nu) &= \gamma_{\mu, i}^D(\omega_J,\mu, \nu),
\\
\nu\frac{d}{d\nu} \ln D_{h/i}(z_h, {\bm b}, \mu, \nu) &= \gamma_{\nu, i}^D(b, \mu),
\eea
where the $\mu$- and $\nu$-anomalous dimensions are given by
\bea
\gamma_{\mu, q}^D(\omega_J,\mu, \nu) =& \frac{\alpha_s}{\pi} C_F \left(2 \ln\left(\frac{\nu}{\omega_J}\right) + \frac{3}{2}\right),
\\
\gamma_{\mu, g}^D(\omega_J,\mu, \nu) =& \frac{\alpha_s}{\pi} C_A \left(2 \ln\left(\frac{\nu}{\omega_J}\right) + \frac{\beta_0}{2C_A}\right),
\\
\gamma_{\nu, q}^D(b,\mu) = & \frac{\alpha_s}{\pi} C_F \ln\left(\frac{\mu^2}{\mu_b^2}\right),
\\
\gamma_{\nu, g}^D(b,\mu) = & \frac{\alpha_s}{\pi} C_A \ln\left(\frac{\mu^2}{\mu_b^2}\right).
\eea

\subsubsection{Matching onto collinear FFs}
After the UV poles are removed via renormalization, the TMDFFs $D_{j/i}(z_h, {\bm b}, \mu, \nu)$ only contain IR poles which have the expected structure $ \sim\, - 1/\epsilon\, P_{ji}(z_h)$. In the perturbative region $1/b \gg \Lambda_{\rm QCD}$, the TMDFFs can be further matched onto the standard collinear FFs $D_{h/i}(z_h, \mu)$. With the help of this matching procedure, the remaining IR poles can be subtracted. The matching relation is given by
\bea
\label{eq:convolution}
D_{h/i}(z_h, {\bm b}, \mu, \nu) = \frac{1}{z_h^2}\int_{z_h}^1 \frac{d\hat z_h}{\hat z_h} \C_{j\leftarrow i}\left(\frac{z_h}{\hat z_h}, {\bm b}, \mu, \nu \right) D_{h/j}(\hat z_h, \mu) \equiv \frac{1}{z_h^2} \C_{j\leftarrow i}\otimes D_{h/j}(z_h, \mu)\,,
\eea
where the matching coefficients are denoted by $\C_{j\leftarrow i}$. The perturbative results at the parton level for the collinear FFs $D_{j/i}(z_h, \mu)$ in the $\overline{\rm MS}$ scheme are given by
\bea
D_{j/i}(z_h, \mu) = \delta_{ij} \delta(1-z_h) + \frac{\alpha_s}{2\pi}\left(-\frac{1}{\epsilon}\right) P_{ji}(z_h) \,.
\eea
Together with the expressions for the TMDFFs in Eq.~\eqref{eq:TMDb}, we find that the matching coefficients in $b$-space are given by
\begin{subequations}
\label{eq:C-tilde}
\bea
\C_{q'\leftarrow q}(z_h, {\bm b}, \mu, \nu) = & 
\delta_{qq'}\Bigg\{\delta(1-z_h)   
- \frac{\alpha_s}{2\pi} \ln\left(\frac{\mu^2}{z_h^2\mu_b^2}\right) P_{qq}(z_h)
\nnu
&+\frac{\alpha_s}{2\pi} C_F \bigg[\ln\left(\frac{\mu^2}{\mu_b^2}\right) \left(2 \ln\left(\frac{\nu}{\omega_J}\right) + \frac{3}{2}\right) \delta(1-z_h) + (1-z_h)\bigg]\Bigg\}\,,
\\[.3cm]
\C_{g\leftarrow q}(z_h, {\bm b}, \mu, \nu) =& \frac{\alpha_s}{2 \pi}\Big[ -\ln\left(\frac{\mu^2}{z_h^2\mu_b^2}\right)P_{gq}(z_h) + C_F  z_h\Big]\,,
\\[.3cm]
\C_{g\leftarrow g}(z_h, {\bm b}, \mu, \nu) =& \delta(1-z_h) - \frac{\alpha_s}{2\pi} \ln\left(\frac{\mu^2}{z_h^2\mu_b^2}\right)P_{gg}(z_h)
\nnu
&+\frac{\alpha_s}{2\pi} C_A \bigg[\ln\left(\frac{\mu^2}{\mu_b^2}\right) \left(2 \ln\left(\frac{\nu}{\omega_J}\right) + \frac{\beta_0}{2C_A}\right) \delta(1-z_h)\bigg]\,,
\\[.3cm]
\C_{q\leftarrow g}(z_h, {\bm b}, \mu, \nu) =&  \frac{\alpha_s}{2\pi}\Big[ -\ln\left(\frac{\mu^2}{z_h^2\mu_b^2}\right) P_{qg}(z_h) + \frac{1}{2} T_F z_h (1-z_h) \Big]\,.
\eea
\end{subequations}

\subsection{Soft functions}
The soft function $S_i({\bm \lambda}_\perp, \mu, \nu R)$ is defined as 
\bea
\label{eq:defS}
S( {\bm \lambda}_\perp, \mu,\nu R ) =& 
\langle 0| {\bar Y}_n  \, 
\delta^2\left({\mathcal P}_\perp^{\in J} - {\bm \lambda}_\perp \right) Y_{\bar{n} }  | X\rangle
\langle  X|{\bar Y}_{\bar n}  Y_n   |0\rangle,
\eea
where $Y_{n(\bar n)}$ denotes the soft Wilson line, and ${\mathcal P}_\perp^{\in J}$ indicates the fact that only the soft radiation inside the jet contributes to the hadron transverse momentum with respect to the jet axis. The soft functions $S_i({\bm \lambda}_\perp, \mu, \nu R)$ also contain rapidity divergences and, thus, the rapidity scale $\nu$ arises. The calculation of the soft functions in the perturbative region $\lambda_\perp\gg \Lambda_{\rm QCD}$ is very similar to the standard global soft function that arises in the processes SIDIS, Drell-Yan, and electron-positron annihilation, except that now we restrict the soft radiation to be inside the jet. The final result up to NLO in momentum space is given by
\bea
S_i({\bm \lambda}_\perp, \mu, \nu R) =& \delta^2({\bm \lambda}_\perp)+ C_i  \,  \frac{\alpha_s}{\pi^2} \, e^{\gamma_E \epsilon}  
 \frac{ \Gamma\left(1+\epsilon + \frac{\eta}{2}\right)}{\Gamma\left(1+\frac{\eta}{2}\right)}
\frac{1}{\mu^2} \left(\frac{\mu^2 }{\lambda_\perp^2 }
\right)^{1+\epsilon + \frac{\eta}{2} }  
\nnu
&\times
\frac{1}{\eta} \left( \frac{ \nu \tan (R/2) }{\mu} \right)^\eta 
\Big[1 + {\cal O}(R^2) \Big],
\eea
where we keep the leading contribution in the limit $R\ll 1$. The color factors are given by $C_i = C_F ~(C_A)$ for $i = q~(g)$, respectively. After taking the Fourier transform to $b$-space, we obtain
\bea
S_i({\bm b}, \mu, \nu R) =& \int d^2{\bm \lambda}_\perp e^{-i{\bm \lambda}_\perp\cdot {\bm b}} S_i({\bm \lambda}_\perp, \mu, \nu R) 
\nnu
= & 1 + \frac{\alpha_s}{2\pi} C_i \bigg[ \frac{2}{\eta} \left( - \frac{1}{\epsilon}-\ln\left(\frac{\mu^2}{\mu_b^2}\right)\right) + \frac{1}{\epsilon^2} -\frac{1}{\epsilon} \ln\left(\frac{\nu^2 \tan^2 (R/2)}{\mu^2}\right)
\nnu
& - \ln\left(\frac{\mu^2}{\mu_b^2}\right) \ln\left(\frac{\nu^2 \tan^2(R/2)}{\mu_b^2}\right) + \frac{1}{2}\ln^2\left(\frac{\mu^2}{\mu_b^2}\right) - \frac{\pi^2}{12} \bigg].
\label{eq:softb}
\eea
Note that the same result was obtained in~\cite{Kang:2017mda} in the context of central subjets. Similar to the renormalization of the TMDFFs discussed above, we subtract the UV poles of the soft functions. The renormalized and bare soft functions $S_i({\bm b}, \mu, \nu R)$ are related by
\bea
S_i({\bm b}, \mu, \nu R) = Z_i^S({\bm  b}, \mu, \nu) S_i^{\rm bare}({\bm b}, \mu, \nu R)\,,
\eea
where the multiplicative renormalization constants $Z_i^S$ are given by
\bea
Z_i^S({\bm  b}, \mu, \nu) = 1 + \frac{\alpha_s}{2\pi} C_i \bigg[ \frac{2}{\eta} \left( - \frac{1}{\epsilon}-\ln\left(\frac{\mu^2}{\mu_b^2}\right)\right) + \frac{1}{\epsilon^2} -\frac{1}{\epsilon} \ln\left(\frac{\nu^2 \tan^2 (R/2)}{\mu^2}\right)\bigg]\,.
\eea
The associated RG and RRG equations are given by
\bea
\mu\frac{d}{d\mu} \ln S_i({\bm b}, \mu, \nu R) =& \gamma_{\mu, i}^S(b, \mu,\nu R)\,,
\\
\nu\frac{d}{d\nu} \ln S_i({\bm b}, \mu, \nu R) =& \gamma_{\nu, i}^S(b, \mu)\,,
\eea 
with the $\mu$- and $\nu$-anomalous dimensions 
\bea
\gamma_{\mu, i}^S(b,\mu,\nu R) =& - \frac{\alpha_s}{\pi} C_i  \ln\left(\frac{\nu^2 \tan^2(R/2)}{\mu_b^2}\right)\,,
\\
\gamma_{\nu, i}^S(b,\mu) =& - \frac{\alpha_s}{\pi} C_i \ln\left(\frac{\mu^2}{\mu_b^2}\right)\,.
\eea

\subsection{Solution of the evolution equations and resummation}
In this section, we provide the details about how to solve the RG and RRG equations derived above for three functions of the siTMDFJFs, i.e. the hard matching functions, the TMDFFs and the soft functions. The resummation of all large logarithms is obtained by the following two step process. First, we evaluate all fixed order results at their natural scales which eliminates all large logarithms. Second, we evolve all three functions from their natural scales to a common scale $\mu\sim p_T$. Effectively, this procedure resums all large logarithms in the fixed-order results derived above. We are going to find that it is numerically more convenient to evolve the TMDFFs and the soft functions to the jet scale $\mu\sim p_T R$ and to combine the result at this scale with the hard matching functions to obtain the siTMDFJFs. Then, we evolve the thus obtained siTMDFJFs from $p_T R\to p_T$ using the RG equations for the combined siTMDFJFs rather than using the RG equations for the three separate functions. We are going to find that the siTMDFJFs satisfy the timelike DGLAP evolution equations like their collinear analogous, the semi-inclusive fragmenting jet functions (siFJFs) as studied in~\cite{Kang:2016ehg}. We show that both approaches for solving the RG equations are equivalent. Besides numerical simplifications, using a combined evolution for the siTMDFJFs also makes the relation to traditional TMDFFs more clear. Before discussing the details of the resummation, we start by introducing the traditional definition of ``proper'' TMDs that allow for a parton model interpretation of TMD sensitive observables.

\subsubsection{Proper TMD definitions}
A crucial feature of the results for the TMDFFs $D_{h/i}$ and the soft functions $S_i$ is that both have rapidity divergences, but their product $D_{h/i} S_i$ is free of rapidity divergences as they exactly cancel. This can be seen clearly from the NLO expressions for $D_{h/i}$ in Eq.~\eqref{eq:TMDb} and $S_i$ in Eq.~\eqref{eq:softb}. The same $1/\eta$ poles appear in both expressions but with opposite signs. Following the usual TMD phenomenology~\cite{Collins:2011zzd,Aybat:2011zv,GarciaEchevarria:2011rb}, we thus define the ``proper'' in-jet TMDFFs ${\cal D}_{h/i}^{R}$ as the product
\bea
{\mathcal D}_{h/i}^{R}(z_h, {\bm b}; \mu) \equiv D_{h/i}(z_h, {\bm b}, \mu, \nu)\, S_i({\bm b}, \mu, \nu R)\,,
\label{eq:newD}
\eea
where the superscript $R$ reminds us that it represents the hadron distribution within a jet of the radius $R$. The cancelation of rapidity divergences for ${\cal D}_{h/i}^{R}$ can be traced back to the fact that the soft radiation is restricted to be only inside the jet. Note that the TMDFFs $D_{h/i}$ for the in-jet calculation turned out to be the same as for other TMD sensitive observables~\cite{Collins:2011zzd,Aybat:2011zv,GarciaEchevarria:2011rb} and they do not depend on $R$, as discussed above. However, the soft functions are different in the sense that the soft radiation is restricted to be only inside the jet. Instead, for the ``global'' soft functions that are relevant for SIDIS and electron-positron annihilation~\cite{Collins:1981uk,Ji:2004wu,Ji:2004xq}, there is no such phase space constraint. The additional phase space restriction encountered here cuts off half of the rapidity divergences compared to the global soft functions. This leads to the cancelation of the rapidity divergences in Eq.~\eqref{eq:newD} for the product ${\mathcal D}_{h/i}^{R} = D_{h/i} S_i$. 

To be more specific, we present results for the global soft functions as well. We use $\hat S_i({\bm b},\mu,\nu)$ to denote the global soft function in Fourier transform space. Without any phase space constraints on the soft radiation, we obtain the following expression for the global soft function in momentum space~\cite{Chiu:2012ir,Kasemets:2015uus}
\bea
{\hat S}_i({\bm \lambda}_\perp, \mu, \nu) =&\, \delta^2({\bm \lambda}_\perp) 
+ C_i  \,  \frac{\alpha_s}{\pi^2} \, e^{\gamma_E \epsilon}  
 \frac{ \Gamma\left(1+\epsilon + \frac{\eta}{2}\right)}{\Gamma\left(1+\frac{\eta}{2}\right)}
\frac{1}{\mu^2} \left(\frac{\mu^2 }{\lambda_\perp^2 }
\right)^{1+\epsilon + \frac{\eta}{2} }  
\nnu
&\times
 \left( \frac{ \nu }{\mu} \right)^\eta 
\frac{2^{-\eta}\Gamma(\frac{1-\eta}{2})\Gamma(\frac{\eta}{2})}{\sqrt{\pi}}\,.
\eea
After taking the Fourier transform to $b$-space as in Eq.~(\ref{eq:softb}) and expanding around $\eta,\epsilon\to 0$, we find
\bea
{\hat S}_i({\bm b}, \mu, \nu) =& \,1 + \frac{\alpha_s}{2\pi} C_i \bigg[ \frac{4}{\eta} \left( - \frac{1}{\epsilon}-\ln\left(\frac{\mu^2}{\mu_b^2}\right)\right) + \frac{2}{\epsilon^2} -\frac{2}{\epsilon} \ln\left(\frac{\nu^2}{\mu^2}\right)
\nnu
& - 2\ln\left(\frac{\mu^2}{\mu_b^2}\right) \ln\left(\frac{\nu^2}{\mu_b^2}\right) + \ln^2\left(\frac{\mu^2}{\mu_b^2}\right) - \frac{\pi^2}{6} \bigg]\,.
\eea
Comparing this result with the $S_i({\bm b},\mu,\nu R)$ in Eq.~(\ref{eq:softb}), we find that the ${\cal O}(\alpha_s)$ terms differ by an overall factor of 2 and $\nu \leftrightarrow \nu\tan(R/2)$. The ``proper'' standard TMDFFs $\hat{\cal D}_{h/i}$ as they appear in SIDIS and electron-positron annihilation are then defined as
\bea
\hat {\mathcal D}_{h/i}(z_h, {\bm b}; \mu) \equiv D_{h/i}(z_h, {\bm b}, \mu, \nu)\sqrt{{\hat S}_i({\bm b}, \mu, \nu)}\,.
\label{eq:standard}
\eea
This product is also free of rapidity divergences allowing for a parton model type interpretation of TMD sensitive observables~\cite{Collins:2011zzd,Aybat:2011zv,GarciaEchevarria:2011rb}. It is important to work out the exact relation between the in-jet TMDFFs ${\mathcal D}_{h/i}^{R}$ considered in this work and the standard TMDFFs $\hat {\mathcal D}_{h/i}$. We will discuss this relation in more detail after deriving the solution of the RG and RRG equations in the next section. 

\subsubsection{Hard matching functions}
 We start with the RG equations for the hard matching functions $\h_{i\to j}$, see Eq.~\eqref{eq:RG_H}. Note that the anomalous dimensions $\gamma_{ij}(z,\omega_J R,\mu)$ in Eq.~\eqref{eq:gamma_H} contain a purely diagonal piece $\delta_{ij}\delta(1-z) \Gamma_i(\omega_J R,\mu)$ and the Altarelli-Parisi splitting functions $P_{ji}(z)$ similar to the timelike DGLAP. We are going to separate these two parts of the anomalous dimensions and the associated evolution. The purely diagonal or non-DGLAP pieces of $\gamma_{ij}(z,\omega_J R,\mu)$ are going to cancel with the respective terms of the anomalous dimensions of the TMDFFs and the soft functions yielding a standard DGLAP evolution equation for the siTMDFJFs. To that extend, we start by writing the functions $\h_{i\to j}$ as
\bea
\label{eq:HC}
\h_{i\to j}(z, \omega_J R, \mu) = {\mathcal E}_i(\omega_J R,\mu) \, {\mathcal C}_{i\to j}(z, \omega_J R, \mu),
\eea
where the ${\mathcal C}_{i\to j}(z, \omega_J R, \mu)$ satisfy evolution equations where the anomalous dimensions are given only by the Altarelli-Parisi splitting functions
\bea
\label{eq:RG_C}
\mu\frac{d}{d\mu} {\mathcal C}_{i\to j}(z, \omega_J R, \mu) =  \frac{\alpha_s}{2\pi}\sum_k \int_z^1 \frac{dz'}{z'} 
P_{ki}\left(\frac{z}{z'}\right) {\mathcal C}_{k\to j}(z', \omega_J R, \mu)\,.
\eea
Note that these evolution equations are similar to DGLAP equations but here we still have four coupled equations. Only the combined siTMDFJFs are going to satisfy the standard timelike DGLAP evolution equations, see Eq.~\eqref{eq:siTMD-DGLAP} below. 

The functions ${\mathcal E}_i(\omega_J R,\mu)$ satisfy multiplicative RG equations
\bea
\label{eq:evo_E}
\mu \frac{d}{d\mu} \ln {\mathcal E}_i(\omega_J R,\mu)  = \Gamma_i(\omega_J R,\mu)\,,
\eea
where the $\Gamma_i(\omega_J R,\mu)$ are given in Eq.~\eqref{eq:Gamma_i}. The solution for the multiplicative RG equations can be written as
\bea
{\mathcal E}_i(\omega_J R,\mu) = {\mathcal E}_i(\omega_J R,\mu_J) \exp\left(\int_{\mu_J}^\mu \frac{d\mu'}{\mu'} \Gamma_i(\omega_J R,\mu')\right)\,.
\eea
The fixed-order results for ${\mathcal E}_i(\omega_J R,\mu)$ can be obtained from Eq.~\eqref{eq:hard} and are given by
\begin{subequations}
\bea
{\mathcal E}_q(\omega_J R,\mu)  &= 1+\frac{\alpha_s}{2\pi} C_F \left(-\frac{L^2}{2} - \frac{3}{2} L \right)\,,
\\
{\mathcal E}_g(\omega_J R,\mu)  &= 1+ \frac{\alpha_s}{2\pi}\left(-C_A\frac{L^2}{2} - \frac{\beta_0}{2} L\right)\,.
\eea 
\end{subequations}
By choosing $\mu_J=p_TR$, we obtain ${\mathcal E}_i(\omega_J R,\mu_J) = 1$ as the initial condition for the evolution in Eq.~\eqref{eq:evo_E}. Using this result in Eq.~\eqref{eq:HC} above, we can write the hard matching functions as
\bea
\label{eq:solution_H}
\h_{i\to j}(z, \omega_J R, \mu)  = \exp\left(\int_{\mu_J}^\mu \frac{d\mu'}{\mu'} \Gamma_i(\omega_J R,\mu')\right) {\mathcal C}_{i\to j}(z, \omega_J R, \mu)\,.
\eea
The functions ${\mathcal C_{i\to j}}(z, \omega_J R, \mu)$ still need to be evolved from $\mu\sim \mu_J = p_T R$ to $\mu\sim p_T$ using the evolution equations in Eq.~\eqref{eq:RG_C} above. Their fixed order expressions are given by
\begin{subequations}
\label{eq:coefficient}
\bea
{\mathcal C}_{q\to q'}(z,\omega_J R,\mu) 
  =& \delta_{qq'}\delta(1-z) +\delta_{qq'} \frac{\alpha_s}{2\pi} \bigg[ 
  C_F \delta(1-z) \frac{\pi^2}{12}+
  P_{qq}(z) L
  \nnu
 & -2C_F(1+z^2)\left(\frac{\ln(1-z)}{1-z}\right)_+ -C_F(1-z)  \bigg] 
\,, \\[.2cm]
 {\mathcal C}_{q\to g}(z,\omega_J R,\mu) 
 =&\frac{\alpha_s}{2\pi}\bigg[\Big(L - 2 \ln(1-z) \Big) P_{gq}(z) - C_Fz \bigg]
\,, \\[.2cm]
{\mathcal C}_{g\to g}(z, \omega_J R, \mu) 
=& \delta(1-z) + \frac{\alpha_s}{2\pi}\bigg[ \delta(1-z) \frac{\pi^2}{12} + P_{gg}(z) L - \frac{4C_A (1-z+z^2)^2}{z} \left(\frac{\ln(1-z)}{1-z}\right)_{+} \bigg]\,, \\[.2cm]
{\mathcal C}_{g\to q}(z,\omega_J R, \mu) 
 =&  \frac{\alpha_s}{2\pi}\bigg[\Big(L - 2\ln(1-z) \Big)  P_{qg}(z) - T_F 2z(1-z) \bigg]
\,.  
\eea
\end{subequations}

\subsubsection{Relation between in-jet and standard TMDFFs}
We are now going to derive the solution of the evolution equations for the TMDFFs and the soft functions in order to obtain the proper in-jet TMDFFs ${\mathcal D}_{h/i}^{R}(z_h, {\bm b}; \mu)$ as defined in Eq.~\eqref{eq:newD}. For comparison, we also show the results for the standard TMDFFs $\hat {\mathcal D}_{h/i}(z_h, {\bm b}; \mu)$ as in Eq.~\eqref{eq:standard}. We start by evolving $D_{h/i}(z_h, {\bm b}, \mu, \nu)$, $S_i({\bm b}, \mu, \nu, R)$ and $\hat S_i({\bm b}, \mu, \nu)$ using their RG and RRG equations. From the perturbative calculations above, we find that the natural scales for the TMDFFs $D_{h/i}$, and the two soft functions $S_i, \hat S_i$ are given by
\begin{subequations}
\label{eq:scale}
\bea
&\mu_D \sim \mu_b, \qquad  \nu_D\sim \omega_J\,,
\\
&\mu_S \sim \mu_b, \qquad \nu_S\sim \frac{\mu_b}{\tan (R/2)}\,,
\\
&\mu_{\hat S} \sim \mu_b, \qquad \nu_{\hat S}\sim \mu_b\,. 
\eea
\end{subequations}
To be consistent with the standard Collins-Soper-Sterman (CSS) formalism~\cite{Collins:1984kg}, we evolve from the natural scales of $D_{h/i}$ and $S_i,~\hat S_i$ as given in Eq.~\eqref{eq:scale}, to a common scale $\mu$ and $\nu$. In terms of the ``proper'' TMDFFs in~\eqref{eq:newD}, the initial conditions for the evolution is given by
\bea
{\mathcal D}_{h/i}^{R}(z_h, {\bm b}; \mu_b) &\equiv D_{h/i}(z_h, {\bm b}, \mu_D, \nu_D) S_i({\bm b}, \mu_S, \nu_S R)\,,
\\
\hat {\mathcal D}_{h/i}(z_h, {\bm b}; \mu_b) & \equiv D_{h/i}(z_h, {\bm b}, \mu_D, \nu_D) \sqrt{\hat S_i({\bm b}, \mu_{\hat S}, \nu_{\hat S})}\,.
\eea
It might be instructive to point out that the ``proper'' TMDs chosen as such are equal perturbatively when evaluated at their natural scales,
\bea
\label{eq:input}
{\mathcal D}_{h/i}^{R}(z_h, {\bm b}; \mu_b) = \hat {\mathcal D}_{h/i}(z_h, {\bm b}; \mu_b)\,.
\eea
This can be directly verified from the perturbative expressions given above. At this point, it might be instructive to point out that according to Eq.~\eqref{eq:scale}, the natural rapidity scales for two soft functions $S_i$ and $\hat S_i$ are quite different, $\nu_S/\nu_{\hat S} = 1/\tan(R/2) \gg 1$ in the small jet radius limit $R\ll 1$. Since the ``proper'' TMDs do not contain rapidity divergences anymore, the $\nu$-dependence will naturally disappear in the end when one evolves to the common rapidity scales. After solving the corresponding RG and RRG equations, we can write the final result in the following form
\bea
\label{eq:temp}
{\mathcal D}_{h/i}^{R}(z_h, {\bm b}; \mu) &= {\mathcal D}_{h/i}^{R}(z_h, {\bm b}; \mu_b) \exp\left[- \int_{\mu_{b}}^\mu\frac{d\mu'}{\mu'}\left(\Gamma_{\rm cusp}^i\ln \left(\frac{\mu_J^2}{\mu'^2}\right) + \gamma^i\right)\right]\,,
\\
\label{eq:temp-stand}
\hat {\mathcal D}_{h/i}(z_h, {\bm b}; \mu) &= \hat {\mathcal D}_{h/i}(z_h, {\bm b}; \mu_b) \exp\left[- \int_{\mu_{b}}^\mu\frac{d\mu'}{\mu'}\left(\Gamma_{\rm cusp}^i\ln \left(\frac{\mu^2}{\mu'^2}\right) + \gamma^i\right)\right]\,.
\eea
Here the cusp anomalous dimension $\Gamma_{\rm cusp}^i$ and the non-cusp $\gamma^i$ allow for a perturbative evaluation as $\Gamma_{\rm cusp}^i = \sum_{n} \Gamma^i_{n-1} \left(\frac{\alpha_s}{\pi}\right)^n$ and 
likewise for $\gamma^i$. The first coefficients can be obtained from our calculation and are given by
\bea
\Gamma^q_{0} & = C_F\,, \quad \quad \gamma^q_{0}  = -\frac{3}{2} C_F\,,
\\
\Gamma^g_{0} & = C_A\,, \quad \quad \gamma^g_{0}  = -\frac{\beta_0}{2}\,.
\eea
The higher-order expressions can be found for example in Ref.~\cite{Echevarria:2012pw}. The ``proper'' in-jet TMDs ${\mathcal D}_{h/i}^{R}$ in Eq.~\eqref{eq:temp} may be further expressed in terms of the ``proper'' standard TMDs $\hat {\mathcal D}_{h/i}$ in Eq.~\eqref{eq:temp-stand} as
\bea
{\mathcal D}_{h/i}^{R}(z_h, {\bm b}; \mu) =&\; {\mathcal D}_{h/i}^{R}(z_h, {\bm b}; \mu_b) \exp\left[ - \int_{\mu_{b}}^{\mu_J}\frac{d\mu'}{\mu'}\left(\Gamma_{\rm cusp}^i\ln \left(\frac{\mu_J^2}{\mu'^2}\right) + \gamma^i\right)\right] 
\nnu
&\times 
\exp\left[-\int_{\mu_J}^{\mu}\frac{d\mu'}{\mu'}\left(\Gamma_{\rm cusp}^i\ln \left(\frac{\mu_J^2}{\mu'^2}\right) + \gamma^i\right)\right]
\nnu
=&\;
\hat {\mathcal D}_{h/i}(z_h, {\bm b}; \mu_J) \exp\left[-\int_{\mu_J}^{\mu}\frac{d\mu'}{\mu'}\left(\Gamma_{\rm cusp}^i\ln \left(\frac{\mu_J^2}{\mu'^2}\right) + \gamma^i\right)\right]\,.
\label{eq:solution_D}
\eea
To obtain the second line we made use of Eq.~\eqref{eq:input}. In other words, the evolved ``proper'' TMDs obtained for the hadron distribution inside jets ${\mathcal D}_{h/i}^{R}(z_h, {\bm b}; \mu)$ at scale $\mu$ is related to the standard TMDs $\hat {\mathcal D}_{h/i}(z_h, {\bm b}; \mu_J)$ evaluated at scale $\mu_J$ multiplied by an overall factor. This overall factor is given by an exponential involving an integration over $\mu'$ from scales $\mu_J$ to $\mu$. Since scales $\mu_J$ and $\mu$ are both in the perturbative regime, $\mu_J,\mu\gg\Lambda_{\rm QCD}$, we find that the relation between the in-jet TMDs ${\cal D}_{h/i}^{R}$ and the standard TMDs ${\hat {\cal D}}_{h/i}$ is purely perturbative.

\subsubsection{Solution for the siTMDFJFs}
We proceed by combining the above results in order to obtain the evolved siTMDFJFs $\GG_{c}^h(z, z_h, \omega_J R, {\bm j}_\perp, \mu)$. Starting from Eq.~\eqref{eq:GG} and by using the relation
\bea
\delta^2\left(z_h {\bm \lambda}_\perp + {\bm k}_\perp - {\bm j}_\perp\right) =\frac{1}{z_h^2} \int \frac{d^2{\bm b}}{(2\pi)^2} \exp\left(-{i\left({\bm \lambda}_\perp + \frac{{\bm k}_\perp}{z_h} - \frac{{\bm j}_\perp}{z_h}\right)\cdot {\bm b}}\right),
\eea
one finds
\bea
\GG_{c}^h(z, z_h, \omega_J R, {\bm j}_\perp, \mu)  =&\; \h_{c\to i}(z, \omega_JR, \mu) \int \frac{d^2{\bm b}}{(2\pi)^2} 
e^{i\,{\bm j}_\perp\cdot {\bm b}/z_h}
D_{h/i}(z_h, {\bm b}, \mu, \nu) S_i({\bm b}, \mu, \nu R)
\nnu
=&\;
\h_{c\to i}(z, \omega_J R, \mu) \int \frac{d^2{\bm b}}{(2\pi)^2} e^{i\,{\bm j}_\perp\cdot {\bm b}/z_h} {\mathcal D}_{h/i}^{R}(z_h, {\bm b}; \mu)\,.
\eea
Now 	we can plug in the result of the hard matching coefficients $\h_{c\to i}(z, \omega_JR, \mu)$ in Eq.~\eqref{eq:solution_H} where we separated and solved the RG equations for the functions ${\cal E}_i(\omega_J R,\mu)$. In addition, we use the results of the evolved in-jet TMDs ${\mathcal D}_{h/i}^{R}(z_h, {\bm b}; \mu)$ in Eq.~\eqref{eq:solution_D}. We find that the siTMDFJFs may eventually be expressed as
\bea
\label{eq:final-pert}
\GG_{c}^h(z, z_h, \omega_J R, {\bm j}_\perp, \mu)  = {\mathcal C}_{c\to i}(z, \omega_JR, \mu) \int \frac{d^2{\bm b}}{(2\pi)^2} e^{i\,{\bm j}_\perp\cdot {\bm b}/z_h} \hat {\mathcal D}_{h/i}(z_h, {\bm b}; \mu_J)\,. 
\eea
It is important to note that here we are able to write the result in terms of the standard TMDs $\hat{\cal D}_{h/i}$. This is possible since the evolution between the scales $\mu_J$ and $\mu$ of ${\cal E}_i(\omega_J R,\mu)$ cancels with the overall multiplicative factor found in Eq.~(\ref{eq:solution_D}) for the in-jet TMDs when written in terms of the standard TMDs. Specifically, we have
\bea
\exp\left(\int_{\mu_J}^\mu \frac{d\mu'}{\mu'} \Gamma_i(\omega_J R, \mu')\right) \exp\left[-\int_{\mu_J}^{\mu}\frac{d\mu'}{\mu'}\left(\Gamma_{\rm cusp}^i\ln \left(\frac{\mu_J^2}{\mu'^2}\right) + \gamma^i\right)\right] = 1\,. 
\eea
The result in Eq.~\eqref{eq:final-pert} constitutes the most important part of our work. It explicitly demonstrates that the hadron transverse momentum distribution within jets is related to the standard TMDFFs (as measured in SIDIS and electron-positron annihilation) probed at the jet scale $\mu_J \sim p_T R$. Eventually, we can write the result as
\bea
\label{eq:final-pert-momentum}
\GG_{c}^h(z, z_h, \omega_J R, {\bm j}_\perp, \mu)  = {\mathcal C}_{c\to i}(z, \omega_JR, \mu)\, \hat {\mathcal D}_{h/i}(z_h, {\bm j}_\perp; \mu_J)\,, 
\eea
where we used the inverse Fourier transform as defined in Eq.~\eqref{eq:inverse-Fourier} to obtain the TMDFFs in momentum space
\bea
\label{eq:TMD-momentum}
\hat {\mathcal D}_{h/i}(z_h, {\bm j}_\perp; \mu_J) = \int \frac{d^2{\bm b}}{(2\pi)^2} e^{i\,{\bm j}_\perp\cdot {\bm b}/z_h} \hat {\mathcal D}_{h/i}(z_h, {\bm b}; \mu_J)\,. 
\eea

\subsection{Final expression for the siTMDFJFs}
\label{subsect:siTMDFJFs}
In the perturbative region where $1/b\gg \Lambda_{\rm QCD}$, one can further match the TMDFFs $\hat {\mathcal D}_{h/i}(z_h, {\bm b}; \mu_b)$ onto the standard collinear FFs $D_{h/i}(z_h, \mu_b)$ as
\bea
\hat {\mathcal D}_{h/i}(z_h, {\bm b}; \mu_b) = \frac{1}{z_h^2} \int_{z_h}^1 \frac{d\hat z_h}{\hat z_h} C_{j\leftarrow i}\left(\frac{z_h}{\hat z_h}, \mu_b\right) D_{h/j}(\hat z_h, \mu_b)\,.
\eea
Using the coefficients $\C_{j\leftarrow i}$ given in Eq.~\eqref{eq:C-tilde} and the perturbative expressions for the soft functions, one obtains
\begin{subequations}\label{eq:matching_C}
\bea
C_{q'\leftarrow q}(z_h, \mu_b) &= \delta_{qq'}\left[\delta(1-z_h) + \frac{\alpha_s}{\pi}\left(-C_F\frac{\pi^2}{24}\delta(1-z_h)
+\frac{C_F}{2} (1-z_h) + P_{qq}(z_h)\ln z_h\right)\right]\,,
\\[.2cm]
C_{g\leftarrow q}(z_h, \mu_b) &= \frac{\alpha_s}{\pi}\left[\frac{C_F}{2} z_h+ P_{gq}(z_h)\ln z_h\right]\,,
\\[.2cm]
C_{g\leftarrow g}(z_h, \mu_b) &= \delta(1-z_h) + \frac{\alpha_s}{\pi}\left[-C_A \frac{\pi^2}{24}\delta(1-z_h) + P_{gg}(z_h)\ln z_h \right]\,,
\\[.2cm]
C_{q\leftarrow g}(z_h, \mu_b) &=  \frac{\alpha_s}{\pi}\Big[T_F z_h(1-z_h)+ P_{qg}(z_h) \ln z_h\Big]\,.
\eea
\end{subequations}
It might be instructive to point out that the above matching coefficients are computed in the standard $\overline{\rm MS}$ scheme, which differs from the simplest minimal subtraction scheme by inserting a factor $S_\epsilon$ for each loop in the counter-terms with $S_\epsilon = (4\pi e^{-\gamma_E})^\epsilon$. However, in the so-called Collins-11 definition of TMDs, this factor was changed to $S_\epsilon^{\rm JCC} = (4\pi)^\epsilon/\Gamma(1-\epsilon)$~\cite{Collins:2011zzd}. We refer to the latter scheme as $\overline{\rm MS}^{\rm JCC}$, in which the $\pi^2$ terms are absent in Eqs.~\eqref{eq:matching_C}(a) and (c). This is compensated for by the fact that there are no $\pi^2$-constants in the expressions for the functions $\mathcal C_{i\to j}$ in Eqs.~\eqref{eq:coefficient}(a) and (c) in the $\overline{\rm MS}^{\rm JCC}$ scheme. 

So far, we have discussed the evolution of the siTMDFJFs in the perturbative region, i.e. for $1/b\gg \Lambda_{\rm QCD}$. It is well-known that the evolution of TMDs contains a non-perturbative component in the large-$b$ region. We treat the large-$b$ region by adopting the usual $b_*$-prescription~\cite{Collins:1984kg}. Alternative approaches can be found in~\cite{Kulesza:2002rh,Qiu:2000hf,Catani:2015vma,Ebert:2016gcn,Monni:2016ktx}. One defines $b_*$ as
\bea
b_* = \frac{b}{\sqrt{1+b^2/b_{\rm max}^2}},
\eea
where the quantity $b_{\rm max}$ is introduced such that $b_*\to b$ at small $b\ll b_{\rm max}$, whereas it approaches the limit $b\to b_{\rm max}$ in the large $b$-region. Using this prescription and the matching coefficients in Eq.~(\ref{eq:matching_C}), we can write the evolved TMDFFs in Eq.~\eqref{eq:TMD-momentum} as
\bea
\hat {\mathcal D}_{h/i}(z_h, {\bm j}_\perp; \mu_J) &= \frac{1}{z_h^2}\int \frac{b\,db}{2\pi} J_0(j_\perp b/z) C_{j\leftarrow i}\otimes D_{h/j}(z_h, \mu_{b_*}) e^{-S_{\rm pert}^i(b_*, \mu_J) - S_{\rm NP}^i(b, \mu_J)}\,.
\eea
Here, $S_{\rm pert}^i(b_*, \mu_J)$ is the perturbative Sudakov factor 
\bea
\label{eq:S-pert}
S_{\rm pert}^i(b_*, \mu_J) = \int_{\mu_{b_*}}^{\mu_J}\frac{d\mu'}{\mu'}\left(\Gamma_{\rm cusp}^i\ln \left(\frac{\mu_J^2}{\mu'^2}\right) + \gamma^i\right)\,,
\eea
and $S_{\rm NP}^i(b, \mu_J)$ is the non-perturbative Sudakov factor. We will discuss them in detail in the phenomenological  Sec.~\ref{sec:pheno} below. 

With all these relevant ingredients available, we may then compute the siTMDFJFs following Eq.~\eqref{eq:final-pert-momentum}. By using the evolution equations for the functions ${\mathcal C}_{i\to j}$ in Eq.~\eqref{eq:RG_C} and the expressions for siTMDFJFs in Eq.~\eqref{eq:final-pert-momentum}, we find that the siTMDFJFs satisfy the standard timelike DGLAP evolution equations
\bea
\label{eq:siTMD-DGLAP}
\mu\frac{d}{d\mu} \GG_{i}^h(z, z_h, \omega_J R, {\bm j}_\perp, \mu) =  \frac{\alpha_s}{2\pi}\sum_j \int_z^1 \frac{dz'}{z'} 
P_{ji}\left(\frac{z}{z'}\right) \GG_{j}^h(z', z_h, \omega_J R, {\bm j}_\perp, \mu)\,. 
\eea
This result for the evolution equations of the siTMDFJFs was to be expected. Following our factorization expression for the differential cross section in Eq.~\eqref{eq:master}, the product $H_{ab}^c\GG_{c}^h$ should be $\mu$-independent order by order analogously to single inclusive hadron production~\cite{Ellis:1985er,Aversa:1988vb}. Thus, it is natural that the $\GG_{c}^h$ follow the same DGLAP evolution equations as those for the usual collinear FFs $D_{h/c}$~\cite{Kang:2016mcy,Kang:2016ehg}. 

We would like to summarize again the following aspects of the evolution structure of the siTMDFJFs. The TMD part of the evolution between $\mu_b$ an $\mu_J$ is governed by the same TMD evolution equations that have been obtained for the standard TMDs as well. The hard matching functions ${\mathcal C}_{c\to i}$ follow RG equations where the anomalous dimensions are given by the usual Altarelli-Parisi splitting functions. The evolution of ${\cal C}_{c\to i}$ is carried out between the jet scale $\mu_J$ and the hard scale $\mu$ allowing for the resummation of logarithms in the jet size parameter $R$. The structure and resummation of single logarithms $\alpha_s^n\ln^n R$ becomes more apparent when combining both contributions to obtain the siTMDFJFs. They follow the standard DGLAP structure as it is usually associated with the resummation of single logarithms in the jet size parameter~\cite{Dasgupta:2014yra,Kang:2016mcy,Kang:2016ehg,Dai:2016hzf}. The obtained structure for the siTMDFJFs provides a convenient method to perform the resummation of all relevant large logarithms. First, we are going to evolve the standard TMDFFs $\hat {\mathcal D}_{h/i}(z_h, {\bm j}_\perp; \mu)$ from the scale $\mu_b$ to $\mu_J$. Second, at the jet scale $\mu_J$ the TMDFFs $\hat {\mathcal D}_{h/i}(z_h, {\bm j}_\perp; \mu_J)$ will be combined with the remaining hard matching functions ${\mathcal C}_{c\to i}(z,\omega_J R,\mu_J)$ in Eq.~\eqref{eq:coefficient} to compute the siTMDFJFs $\GG_{c}^h(z, z_h, \omega_J R, {\bm j}_\perp, \mu_J)$ as
\bea
\GG_{c}^h(z, z_h, \omega_J R, {\bm j}_\perp, \mu_J)  = {\mathcal C}_{c\to i}(z, \omega_JR, \mu_J) \hat {\mathcal D}_{h/i}(z_h, {\bm j}_\perp; \mu_J)\,. 
\eea
Then, we use the DGLAP evolution equations for the siTMDFJFs in Eq.~\eqref{eq:siTMD-DGLAP} to evolve $\GG_{c}^h$ from the scale $\mu_J$ to $\mu$ and, thus, resum logarithms in the jet size parameter $R$.

\section{Phenomenology for $pp\to ({\rm jet} h)X$}
\label{sec:pheno}

In this section, we present numerical results for the transverse momentum distribution of hadrons inside jets for LHC kinematics. We consider an inclusive jet sample $pp\to\text{jet}+ X$ where a hadron $h$ is identified inside the reconstructed jet. Following~\cite{Kaufmann:2015hma,Kang:2016mcy,Kang:2016ehg,Neill:2016vbi}, the factorization theorem for the process $pp\to (\mathrm{jet}h)X$ can be written as
\bea
\label{eq:sigjethX}
\frac{d\sigma^{pp\to (\mathrm{jet}h)X}}{dp_Td\eta dz_h d^2 {\bm j}_\perp}  = & \sum_{a,b,c}\int_{x_a^{\mathrm{min}}}^1\frac{dx_a}{x_a}f_a(x_a,\mu)\int_{x_b^{\mathrm{min}}}^1\frac{dx_b}{x_b} f_b(x_b,\mu)  
\nnu
&\times \,\int^1_{z^{\mathrm{min}}} \frac{dz}{z^2} H^c_{ab}(\hat s,\hat p_T,\hat \eta,\mu)\;
\GG_{c}^h(z, z_h, \omega_J R, {\bm j}_\perp, \mu) \; ,
\eea
where $f_a$ and $f_b$ denote the parton distribution functions (PDFs) in the proton with the corresponding momentum fraction $x_a$ and $x_b$, respectively. For all numerical calculations in this work, we choose the CT14 NLO set of PDFs~\cite{Dulat:2015mca}. The hard functions $H^c_{ab}$ describe the production of an energetic parton $c$ in the hard-scattering event. They have been calculated analytically up to NLO in~\cite{Aversa:1988vb,Jager:2002xm}. The variables $\hat s$, $\hat p_T$ and $\hat\eta$ denote the partonic CM energy, and the transverse momentum and rapidity of parton $c$, respectively. They are related to their hadronic analogues as
\be
\hat s=x_a x_b s,\quad \hat p_T=p_T/z,\quad \hat\eta=\eta-\ln(x_a/x_b)/2\,,
\ee
where $z$ is the momentum fraction transferred from parton $c$ to the observed jet. The lower integration limits $x_a^{\text{min}}$, $x_b^{\text{min}}$ and $z^{\text{min}}$ can be found for example in~\cite{Kaufmann:2015hma,Kang:2016ehg}. Finally, the functions $\GG_{c}^h(z_c, z_h, \omega_J R, {\bm j}_\perp, \mu)$ in Eq.~(\ref{eq:sigjethX}) are the siTMDFJFs as discussed in Section~\ref{sec:fac}. We would like to stress that the cross section does not depend on TMDPDFs but only on the standard collinear PDFs. Unlike the TMDPDFs, collinear PDFs are very well constrained by data and have been determined in global fits in the literature. Therefore, different than for SIDIS, the hadron in-jet fragmentation considered in this work provides an opportunity to disentangle the effects of TMDPDFs and TMDFFs. 

In order to perform numerical calculations, we have to parameterize the non-perturbative Sudakov factors for both quark and gluon TMDFFs. Unfortunately, the quark TMDFFs are not very well constrained so far. The main information for the extraction of quark TMDFFs are obtained from multiplicity distributions of hadrons measured in SIDIS from both the HERMES~\cite{Airapetian:2012ki} and COMPASS~\cite{Adolph:2013stb} experiments. These measurements were performed at relative low momentum scales, with photon virtualities $Q^2$ of several GeV$^2$. Thus, there are potential problems when interpreting the data in terms of the usual leading-twist TMD factorization formalism~\cite{Boglione:2016bph}. In addition, since the factorization for SIDIS involves a convolution of TMDPDFs and TMDFFs, the unambiguous extraction of both functions separately is not straightforward. Therefore, current extractions of quark TMDFFs are subject to large uncertainties. Keeping in mind the remaining large uncertainties in our calculation, we are nevertheless going to present numerical estimates for the hadron transverse momentum distribution within jets and compare to LHC measurements. We choose to use the following parametrization of the non-perturbative Sudakov factor following~\cite{Kang:2015msa,Su:2014wpa}
\bea
S_{\rm NP}^q(b, \mu_J) = \frac{g_2}{2} \ln\left(\frac{b}{b_*}\right) \ln\left(\frac{\mu_J}{Q_0}\right) + \frac{g_h}{z_h^2} b^2,
\eea
with $Q_0^2 = 2.4$ GeV$^2$, $b_{\rm max}=1.5$ GeV$^{-1}$, $g_2 = 0.84$, and $g_h=0.042$. Other parametrizations for the non-perturbative Sudakov factor for TMDFFs have been discussed in~\cite{Bacchetta:2017gcc,Echevarria:2014xaa}.

Furthermore, we note that the non-perturbative Sudakov factor for the gluon TMDFF is not constrained at all so far. For our numerical calculations, we are going to follow \cite{Balazs:1997hv,Balazs:2000wv,Balazs:2007hr} and adopt a parameterization of the gluon non-perturbative Sudakov factor similar to that for quarks as
\bea
S_{\rm NP}^g(b, \mu_J) = \frac{C_A}{C_F} \frac{g_2}{2} \ln\left(\frac{b}{b_*}\right) \ln\left(\frac{\mu_J}{Q_0}\right) + \frac{g_h}{z_h^2} b^2\,.
\eea
In comparison to the quark parametrization, the coefficient of the term $\sim\ln \mu_J$ is enhanced by a color factor $C_A/C_F$, whereas the intrinsic part $\sim g_h$ is kept unchanged. 
\bef
\includegraphics[width=2.6in]{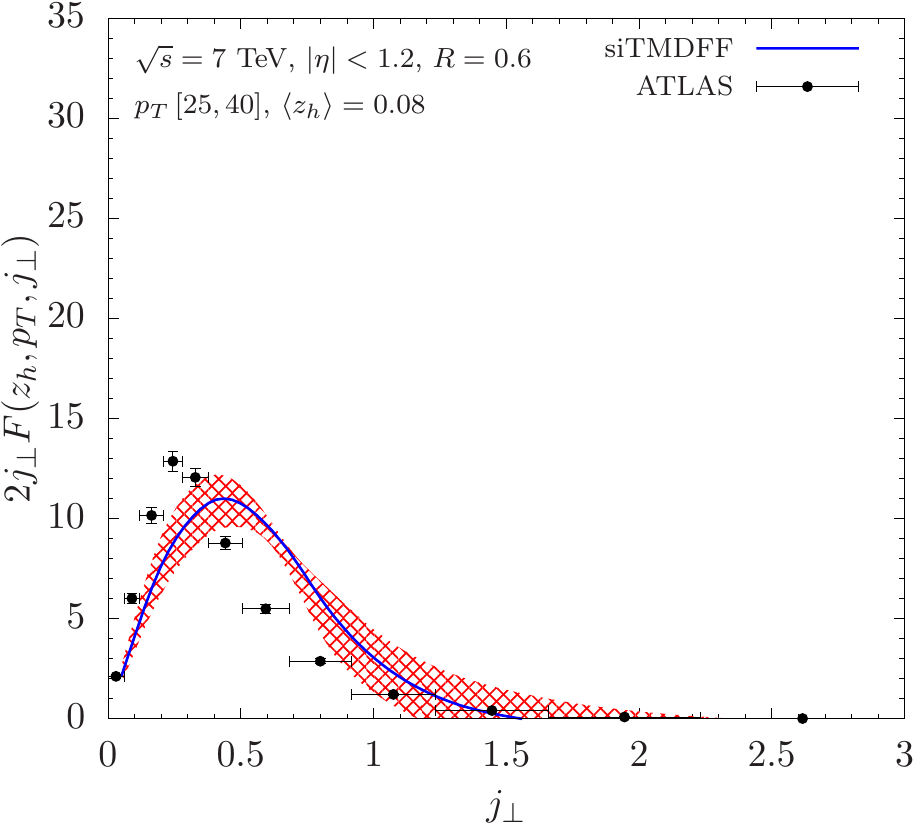}
\hskip 0.3in
\includegraphics[width=2.6in]{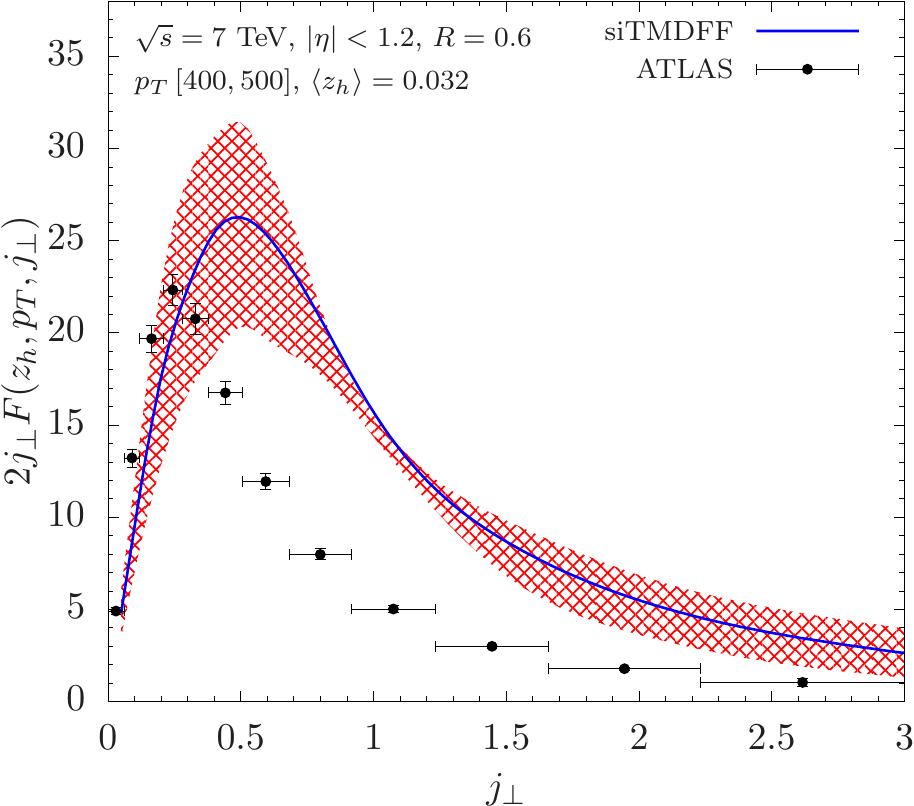}
\caption{Hadron $j_\perp$-distributions within jets in $pp$ collisions at $\sqrt{s} = 7$ TeV. Jets are taken into account in the rapidity interval $|\eta| < 1.2$ and they are reconstructed using the \kt~jet algorithm with $R=0.6$. We choose jet transverse momentum bins $25 < p_T < 40$ GeV (left) and $400 < p_T < 500$ GeV (right). The average values $\langle z_h\rangle$ are provided by the experiment, $\langle z_h\rangle = 0.08$ (left) and $0.03$ (right). The uncertainty band is calculated by varying the scales $\mu$ and $\mu_J$ independently by a factor of two around their default values $\mu=p_T$ and $\mu_J = p_T R$, and taking the envelope of these variations.}\label{fig:data} 
\eef

In addition, we use the coefficients $C_{j\leftarrow i}$ in Eq.~\eqref{eq:matching_C} up to the order of $\alpha_s$ for the TMDFFs, and keep $\Gamma^i_{0,1}$ and $\gamma^i_{0}$ in the perturbative Sudakov factor in Eq.~\eqref{eq:S-pert}, which is often referred to as the next-to-leading-logarithm prime (NLL$'$) accuracy. For both $C_{j\leftarrow i}$ in Eq.~\eqref{eq:matching_C} and ${\mathcal C}_{i\to j}$ in Eq.~\eqref{eq:coefficient}, we use the expressions in the $\overline{\rm MS}^{\rm JCC}$ scheme as explained in Sec.~\ref{subsect:siTMDFJFs}, since the TMDFFs that we use for our numerical studies were extracted within this scheme~\cite{Su:2014wpa,Kang:2015msa}. We use the DSS07 parametrization of collinear fragmentation functions for light charged hadrons~\cite{deFlorian:2007ekg}. Together with the choices for the relevant non-perturbative inputs above for both quark and gluon TMDFFs, we are now going to present first numerical estimates for the transverse momentum distribution of hadrons inside jets and compare to the data provided by the ATLAS collaboration~\cite{Aad:2011sc}. We choose the following jet kinematics consistent with the available data at a CM energy of $\sqrt{s}=7$ TeV. The jets are reconstructed using the \kt~algorithm with jet size parameter $R=0.6$, and the jet rapidity is integrated over $|\eta| < 1.2$. The detailed numerical implementation is very similar to the longitudinal momentum distribution of hadrons inside jets~\cite{Kang:2016ehg}, using several numerical techniques developed in the literature~\cite{Vogt:2004ns,Anderle:2015lqa,Bodwin:2014gia}. The RG evolution of the various parts of the cross section is performed as outlined at the end of the last section.

In Fig.~\ref{fig:data}, we present the comparison of our numerical results and the LHC data for the hadron $j_\perp$-distribution inside jets. We make the default scale choices of $\mu = p_T$ and $\mu_J = p_T R$. We explore the scale uncertainty by varying $\mu$ and $\mu_J$ independently by a factor of two around their default values and by taking the envelope of these variations. As an example, we choose the jet transverse momentum bins $25 < p_T < 40$ GeV (left) and $400 < p_T < 500$ GeV (right). The experimental data are presented for the $z_h$-integrated hadron distribution, i.e. with $z_h$ integrated from 0 to 1. This fact hinders a more direct and transparent comparison of our results with the data, since the collinear FFs are only constrained in a finite region $z_h^{\rm min} < z_h < 1$ with $z_h^{\rm min} \gtrsim 0.05$~\cite{deFlorian:2007ekg,deFlorian:2014xna}. Any $z_h< z_h^{\rm min}$ is not constrained and can only be obtained by extrapolation. We choose the value for $z_h$ in our calculations as the average value $\langle z_h\rangle$ that are provided in the experimental publication~\cite{Aad:2011sc}, with $\langle z_h\rangle = 0.08$ and $0.03$ for $25 < p_T < 40$ GeV and $400 < p_T < 500$ GeV, respectively. With this caveat in mind, we find that our calculations based on TMDFFs extracted in the literature at low energy scales of several GeV give a reasonable description of the experimental data. The height of the peak is roughly consistent with the data but our results have a broader $j_\perp$-distribution than the experimental data. We note that at low jet $p_T$ our current numerical estimates agrees somewhat better with the data than in the high $p_T$ region.

\bef 
\includegraphics[width=2.6in]{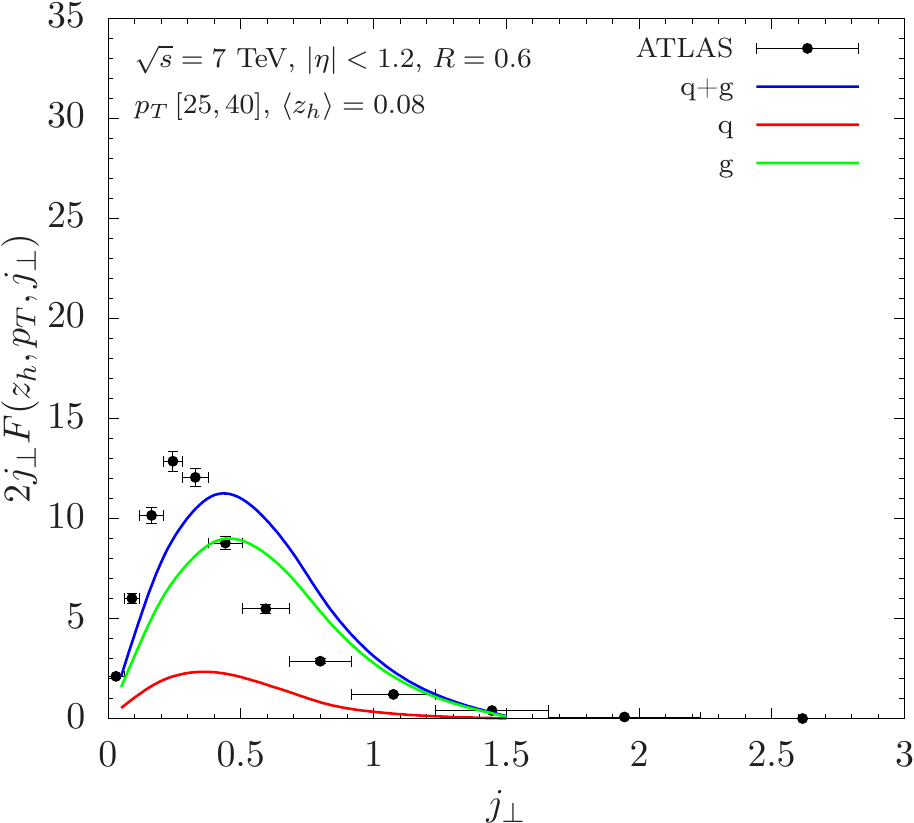}
\hskip 0.3in
\includegraphics[width=2.6in]{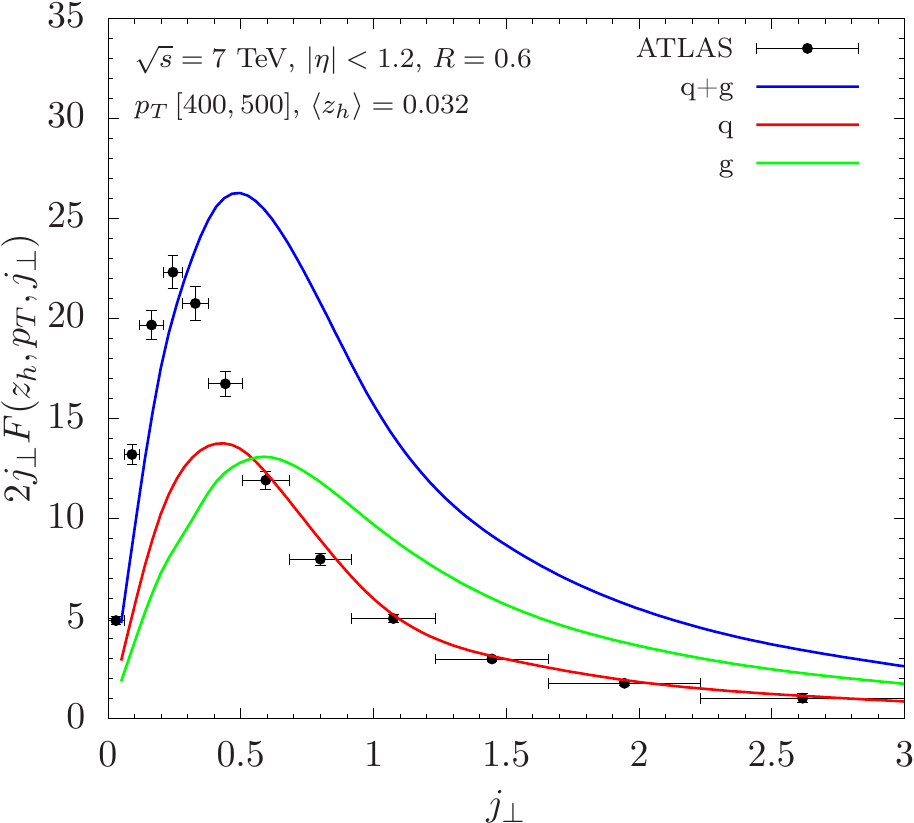}
\caption{Breakdown of the hadron $j_\perp$-distributions inside jets (blue) into quark initiated (red) and gluon initiated (green) TMDFF channels.}\label{fig:qg}
\eef

In Fig.~\ref{fig:qg}, we separate the hadron $j_\perp$-distribution into quark and the gluon TMDFF components. This separation is valid in the TMD region. We find that at lower jet $p_T$, as shown in the left panel of Fig.~\ref{fig:qg}, the gluon channel dominates over the quark channel due to the overwhelmingly abundant $gg$ initiated events in the $pp$ collisions.  The quark TMD fragmenting contribution is suppressed in this region. Therefore, the low jet $p_T$ region provides a ``golden channel'' to extract the gluon TMDFF. At large jet $p_T$ (right panel of Fig.~\ref{fig:qg}), where $qg$ initiated events start to dominate, the quark and the gluon TMDFF contributions therefore become comparable to each other. However, due to the difference in the color charges carried by quarks and gluons, the quark TMD fragmenting process peaks at smaller $j_\perp$. Away from the peak region, the quark contribution drops more dramatically and exhibits a relatively narrow spectrum compared with the gluon contribution. Therefore, the region away from the peak of the $j_\perp$-spectrum will generally be more sensitive to the gluon TMDFF. 

To conclude this section, we provide further discussions of our numerical estimates. First, we would like to emphasize that at the moment we only concentrate on the TMD region and we present numerical results without matching onto NLO fixed-order calculations. In other words, we have not considered the effect of the so-called $Y$-term which can also affect the low $j_\perp$-region, as advocated recently in~\cite{Collins:2016hqq,Boglione:2014oea}. Second, in the TMD region, the non-perturbative parts of the quark TMDFFs have large uncertainties as they have only been constrained from SIDIS data so far, while the gluon TMDFF has not been extracted at all. Third, so far we did not take into account the effect of non-global logarithms (NGLs)~\cite{Dasgupta:2001sh,Banfi:2002hw}. They first arise at next-to-next-to leading order due to the hierarchies caused by different constraints in different phase space regions, and affect our results at the current logarithmic accuracy we are considering. The factorization and resummation of NGLs have been studied recently in great detail, see for example Refs.~\cite{Dasgupta:2012hg,Larkoski:2015zka,Larkoski:2016zzc,Neill:2016stq,Becher:2016mmh,Caron-Huot:2015bja,Hagiwara:2015bia}. We expect to obtain significant improvements of our results in the comparison with the experimental data once all these additional factors are taken into account. A dedicated study including all additional effects will be presented in a forthcoming publication. 

\section{Conclusions}
\label{sec:sum}
In this work, 
we have studied the hadron transverse momentum $j_\perp$-distribution within jets, where $j_\perp$ is defined with respect to the standard jet axis. We set up a factorization formalism that allows for systematic studies of this distribution. As a first step we calculated all the components of the factorization theorem to NLO, and we further resummed all the associated large logarithms $\ln R$ and $\ln(p_T R/j_\perp)$. 
We demonstrated the universality of the TMDFFs that arise for this jet substructure observable and the traditional TMDFFs probed in SIDIS and electron-positron annihilation. We further showed that the hadron distribution within jets produced in $pp$ collisions provides a unique opportunity to study the TMDFFs, especially the gluon TMDFF. For SIDIS and electron-positron annihilation, the gluon TMDFF is usually difficult to access. More specifically, we showed that different than for SIDIS, the $j_\perp$ spectrum within jets only depends on TMDFFs. There is no dependence on TMDPDFs which allows for a more direct extraction of TMDFFs. Furthermore, we found that at LHC energies we are able to control the sensitivity to different TMDFFs by selecting different values of the jet $p_T$. We observed that the low jet $p_T$ region is the ideal region to extract the gluon TMDFF. For large jet $p_T$, the region away from the peak of the $j_\perp$-spectrum can also be sensitive to the gluon TMDFF. 

In the future, several extensions of this work are possible. For instance, in order to extend our calculations to the region where ${j}_\perp \sim p_T R$, we need to match the resummed result onto fixed order calculations. Such a matching calculation includes the full NLO corrections to this spectrum which may also affect the TMD region. In addition, it will be important to study the numerical impact of NGLs. Besides improvements of the perturbative calculation, a more careful study of the non-perturbative Sudakov evolution will be necessary to determine whether the agreement with the data in the region $j_\perp < 1\, {\rm GeV}$ can be improved. Also given the relative simple structure of the TMDFFs and the soft functions considered here, a next-to-next-to leading order calculation is possible which will further push forward the accuracy of the theoretical predictions. In this work, we only considered the phenomenology of $pp$ collisions but the formalism developed here is also directly applicable to $ep$ scattering relevant for a future Electron-Ion Collider (EIC). Other phenomenological studies may include for example a global fit of TMDFFs using also data from SIDIS and electron-positron annihilation. Finally, we are also planning to extend our formalism to the polarized case which is crucial to probe the Collins function, Hyperon polarization inside jets and other types of jet substructure observables.

\acknowledgments

We thank H.-n.~Li, Y.~Makris, A.~Metz, D.~Neill, J.~Qiu, I.~Scimemi, I.~Stewart, and W.~Waalewijn for discussions and comments. This work is supported by the Department of Energy under Contract Nos.~DE-AC52-06NA25396, DE-AC02-05CH11231, DE-FG02-91ER40684 and DE-AC02-06CH11357, and by the Laboratory Directed Research and Development Program of Lawrence Berkeley National Laboratory. X.L. is grateful for the hospitality of the Theoretical Division at Los Alamos National Laboratory and Kavli Institute of Theoretical Physics in Santa Barbara during the completion of this manuscript. 
 
\bibliographystyle{JHEP}
\bibliography{bibliography}

\providecommand{\href}[2]{#2}\begingroup\raggedright\begin{thebibliography}{10}

\bibitem{Altheimer:2013yza}
A.~Altheimer et~al., {\it {Boosted objects and jet substructure at the LHC.
  Report of BOOST2012, held at IFIC Valencia, 23rd-27th of July 2012}},  {\em
  Eur. Phys. J.} {\bf C74} (2014), no.~3 2792,
  [\href{http://arxiv.org/abs/1311.2708}{{\tt arXiv:1311.2708}}].

\bibitem{Adams:2015hiv}
D.~Adams et~al., {\it {Towards an Understanding of the Correlations in Jet
  Substructure}},  {\em Eur. Phys. J.} {\bf C75} (2015), no.~9 409,
  [\href{http://arxiv.org/abs/1504.00679}{{\tt arXiv:1504.00679}}].

\bibitem{Aad:2011sc}
{\bf ATLAS} Collaboration, G.~Aad et~al., {\it {Measurement of the jet
  fragmentation function and transverse profile in proton-proton collisions at
  a center-of-mass energy of 7 TeV with the ATLAS detector}},  {\em Eur. Phys.
  J.} {\bf C71} (2011) 1795, [\href{http://arxiv.org/abs/1109.5816}{{\tt
  arXiv:1109.5816}}].

\bibitem{Aschenauer:2013woa}
E.~C. Aschenauer et~al., {\it {The RHIC Spin Program: Achievements and Future
  Opportunities}},  \href{http://arxiv.org/abs/1304.0079}{{\tt
  arXiv:1304.0079}}.

\bibitem{Aschenauer:2015eha}
E.-C. Aschenauer et~al., {\it {The RHIC SPIN Program: Achievements and Future
  Opportunities}},  \href{http://arxiv.org/abs/1501.01220}{{\tt
  arXiv:1501.01220}}.

\bibitem{Aschenauer:2016our}
E.-C. Aschenauer et~al., {\it {The RHIC Cold QCD Plan for 2017 to 2023: A
  Portal to the EIC}},  \href{http://arxiv.org/abs/1602.03922}{{\tt
  arXiv:1602.03922}}.

\bibitem{Yuan:2007nd}
F.~Yuan, {\it {Azimuthal asymmetric distribution of hadrons inside a jet at
  hadron collider}},  {\em Phys. Rev. Lett.} {\bf 100} (2008) 032003,
  [\href{http://arxiv.org/abs/0709.3272}{{\tt arXiv:0709.3272}}].

\bibitem{Collins:1992kk}
J.~C. Collins, {\it {Fragmentation of transversely polarized quarks probed in
  transverse momentum distributions}},  {\em Nucl. Phys.} {\bf B396} (1993)
  161--182, [\href{http://arxiv.org/abs/hep-ph/9208213}{{\tt hep-ph/9208213}}].

\bibitem{Bain:2016rrv}
R.~Bain, Y.~Makris, and T.~Mehen, {\it {Transverse Momentum Dependent
  Fragmenting Jet Functions with Applications to Quarkonium Production}},  {\em
  JHEP} {\bf 11} (2016) 144, [\href{http://arxiv.org/abs/1610.06508}{{\tt
  arXiv:1610.06508}}].

\bibitem{Neill:2016vbi}
D.~Neill, I.~Scimemi, and W.~J. Waalewijn, {\it {Jet axes and universal
  transverse-momentum-dependent fragmentation}},  {\em JHEP} {\bf 04} (2017)
  020, [\href{http://arxiv.org/abs/1612.04817}{{\tt arXiv:1612.04817}}].

\bibitem{Procura:2009vm}
M.~Procura and I.~W. Stewart, {\it {Quark Fragmentation within an Identified
  Jet}},  {\em Phys. Rev.} {\bf D81} (2010) 074009,
  [\href{http://arxiv.org/abs/0911.4980}{{\tt arXiv:0911.4980}}]. [Erratum:
  Phys. Rev.D83,039902(2011)].

\bibitem{Jain:2011xz}
A.~Jain, M.~Procura, and W.~J. Waalewijn, {\it {Parton Fragmentation within an
  Identified Jet at NNLL}},  {\em JHEP} {\bf 05} (2011) 035,
  [\href{http://arxiv.org/abs/1101.4953}{{\tt arXiv:1101.4953}}].

\bibitem{Arleo:2013tya}
F.~Arleo, M.~Fontannaz, J.-P. Guillet, and C.~L. Nguyen, {\it {Probing
  fragmentation functions from same-side hadron-jet momentum correlations in
  p-p collisions}},  {\em JHEP} {\bf 04} (2014) 147,
  [\href{http://arxiv.org/abs/1311.7356}{{\tt arXiv:1311.7356}}].

\bibitem{Kaufmann:2015hma}
T.~Kaufmann, A.~Mukherjee, and W.~Vogelsang, {\it {Hadron Fragmentation Inside
  Jets in Hadronic Collisions}},  {\em Phys. Rev.} {\bf D92} (2015), no.~5
  054015, [\href{http://arxiv.org/abs/1506.01415}{{\tt arXiv:1506.01415}}].

\bibitem{Chien:2015ctp}
Y.-T. Chien, Z.-B. Kang, F.~Ringer, I.~Vitev, and H.~Xing, {\it {Jet
  fragmentation functions in proton-proton collisions using soft-collinear
  effective theory}},  {\em JHEP} {\bf 05} (2016) 125,
  [\href{http://arxiv.org/abs/1512.06851}{{\tt arXiv:1512.06851}}].

\bibitem{Kang:2016ehg}
Z.-B. Kang, F.~Ringer, and I.~Vitev, {\it {Jet substructure using
  semi-inclusive jet functions in SCET}},  {\em JHEP} {\bf 11} (2016) 155,
  [\href{http://arxiv.org/abs/1606.07063}{{\tt arXiv:1606.07063}}].

\bibitem{Bauer:2000ew}
C.~W. Bauer, S.~Fleming, and M.~E. Luke, {\it {Summing Sudakov logarithms in B
  ---> X(s gamma) in effective field theory}},  {\em Phys. Rev.} {\bf D63}
  (2000) 014006, [\href{http://arxiv.org/abs/hep-ph/0005275}{{\tt
  hep-ph/0005275}}].

\bibitem{Bauer:2000yr}
C.~W. Bauer, S.~Fleming, D.~Pirjol, and I.~W. Stewart, {\it {An Effective field
  theory for collinear and soft gluons: Heavy to light decays}},  {\em Phys.
  Rev.} {\bf D63} (2001) 114020,
  [\href{http://arxiv.org/abs/hep-ph/0011336}{{\tt hep-ph/0011336}}].

\bibitem{Bauer:2001ct}
C.~W. Bauer and I.~W. Stewart, {\it {Invariant operators in collinear effective
  theory}},  {\em Phys. Lett.} {\bf B516} (2001) 134--142,
  [\href{http://arxiv.org/abs/hep-ph/0107001}{{\tt hep-ph/0107001}}].

\bibitem{Bauer:2001yt}
C.~W. Bauer, D.~Pirjol, and I.~W. Stewart, {\it {Soft collinear factorization
  in effective field theory}},  {\em Phys. Rev.} {\bf D65} (2002) 054022,
  [\href{http://arxiv.org/abs/hep-ph/0109045}{{\tt hep-ph/0109045}}].

\bibitem{Bauer:2002nz}
C.~W. Bauer, S.~Fleming, D.~Pirjol, I.~Z. Rothstein, and I.~W. Stewart, {\it
  {Hard scattering factorization from effective field theory}},  {\em Phys.
  Rev.} {\bf D66} (2002) 014017,
  [\href{http://arxiv.org/abs/hep-ph/0202088}{{\tt hep-ph/0202088}}].

\bibitem{Kang:2016mcy}
Z.-B. Kang, F.~Ringer, and I.~Vitev, {\it {The semi-inclusive jet function in
  SCET and small radius resummation for inclusive jet production}},  {\em JHEP}
  {\bf 10} (2016) 125, [\href{http://arxiv.org/abs/1606.06732}{{\tt
  arXiv:1606.06732}}].

\bibitem{Mukherjee:2012uz}
A.~Mukherjee and W.~Vogelsang, {\it {Jet production in (un)polarized pp
  collisions: dependence on jet algorithm}},  {\em Phys. Rev.} {\bf D86} (2012)
  094009, [\href{http://arxiv.org/abs/1209.1785}{{\tt arXiv:1209.1785}}].

\bibitem{Kang:2017mda}
Z.-B. Kang, F.~Ringer, and W.~J. Waalewijn, {\it {The Energy Distribution of
  Subjets and the Jet Shape}},  \href{http://arxiv.org/abs/1705.05375}{{\tt
  arXiv:1705.05375}}.

\bibitem{Jain:2011iu}
A.~Jain, M.~Procura, and W.~J. Waalewijn, {\it {Fully-Unintegrated Parton
  Distribution and Fragmentation Functions at Perturbative $k_T$}},  {\em JHEP}
  {\bf 04} (2012) 132, [\href{http://arxiv.org/abs/1110.0839}{{\tt
  arXiv:1110.0839}}].

\bibitem{Collins:2011zzd}
J.~Collins, {\em {Foundations of perturbative QCD}}.
\newblock Cambridge University Press, 2013.

\bibitem{Becher:2015hka}
T.~Becher, M.~Neubert, L.~Rothen, and D.~Y. Shao, {\it {Effective Field Theory
  for Jet Processes}},  {\em Phys. Rev. Lett.} {\bf 116} (2016), no.~19 192001,
  [\href{http://arxiv.org/abs/1508.06645}{{\tt arXiv:1508.06645}}].

\bibitem{Chien:2015cka}
Y.-T. Chien, A.~Hornig, and C.~Lee, {\it {A Soft-Collinear Mode for Jet Cross
  Sections in Soft Collinear Effective Theory}},
  \href{http://arxiv.org/abs/1509.04287}{{\tt arXiv:1509.04287}}.

\bibitem{Chiu:2012ir}
J.-Y. Chiu, A.~Jain, D.~Neill, and I.~Z. Rothstein, {\it {A Formalism for the
  Systematic Treatment of Rapidity Logarithms in Quantum Field Theory}},  {\em
  JHEP} {\bf 05} (2012) 084, [\href{http://arxiv.org/abs/1202.0814}{{\tt
  arXiv:1202.0814}}].

\bibitem{Echevarria:2014xaa}
M.~G. Echevarria, A.~Idilbi, Z.-B. Kang, and I.~Vitev, {\it {QCD Evolution of
  the Sivers Asymmetry}},  {\em Phys. Rev.} {\bf D89} (2014) 074013,
  [\href{http://arxiv.org/abs/1401.5078}{{\tt arXiv:1401.5078}}].

\bibitem{Aybat:2011zv}
S.~M. Aybat and T.~C. Rogers, {\it {TMD Parton Distribution and Fragmentation
  Functions with QCD Evolution}},  {\em Phys. Rev.} {\bf D83} (2011) 114042,
  [\href{http://arxiv.org/abs/1101.5057}{{\tt arXiv:1101.5057}}].

\bibitem{GarciaEchevarria:2011rb}
M.~G. Echevarria, A.~Idilbi, and I.~Scimemi, {\it {Factorization Theorem For
  Drell-Yan At Low $q_T$ And Transverse Momentum Distributions
  On-The-Light-Cone}},  {\em JHEP} {\bf 07} (2012) 002,
  [\href{http://arxiv.org/abs/1111.4996}{{\tt arXiv:1111.4996}}].

\bibitem{Collins:1981uk}
J.~C. Collins and D.~E. Soper, {\it {Back-To-Back Jets in QCD}},  {\em Nucl.
  Phys.} {\bf B193} (1981) 381. [Erratum: Nucl. Phys.B213,545(1983)].

\bibitem{Ji:2004wu}
X.-d. Ji, J.-p. Ma, and F.~Yuan, {\it {QCD factorization for semi-inclusive
  deep-inelastic scattering at low transverse momentum}},  {\em Phys. Rev.}
  {\bf D71} (2005) 034005, [\href{http://arxiv.org/abs/hep-ph/0404183}{{\tt
  hep-ph/0404183}}].

\bibitem{Ji:2004xq}
X.-d. Ji, J.-P. Ma, and F.~Yuan, {\it {QCD factorization for spin-dependent
  cross sections in DIS and Drell-Yan processes at low transverse momentum}},
  {\em Phys. Lett.} {\bf B597} (2004) 299--308,
  [\href{http://arxiv.org/abs/hep-ph/0405085}{{\tt hep-ph/0405085}}].

\bibitem{Kasemets:2015uus}
T.~Kasemets, W.~J. Waalewijn, and L.~Zeune, {\it {Calculating Soft Radiation at
  One Loop}},  {\em JHEP} {\bf 03} (2016) 153,
  [\href{http://arxiv.org/abs/1512.00857}{{\tt arXiv:1512.00857}}].

\bibitem{Collins:1984kg}
J.~C. Collins, D.~E. Soper, and G.~F. Sterman, {\it {Transverse Momentum
  Distribution in Drell-Yan Pair and W and Z Boson Production}},  {\em Nucl.
  Phys.} {\bf B250} (1985) 199--224.

\bibitem{Echevarria:2012pw}
M.~G. Echevarria, A.~Idilbi, A.~Schafer, and I.~Scimemi, {\it
  {Model-Independent Evolution of Transverse Momentum Dependent Distribution
  Functions (TMDs) at NNLL}},  {\em Eur. Phys. J.} {\bf C73} (2013), no.~12
  2636, [\href{http://arxiv.org/abs/1208.1281}{{\tt arXiv:1208.1281}}].

\bibitem{Kulesza:2002rh}
A.~Kulesza, G.~F. Sterman, and W.~Vogelsang, {\it {Joint resummation in
  electroweak boson production}},  {\em Phys. Rev.} {\bf D66} (2002) 014011,
  [\href{http://arxiv.org/abs/hep-ph/0202251}{{\tt hep-ph/0202251}}].

\bibitem{Qiu:2000hf}
J.-w. Qiu and X.-f. Zhang, {\it {Role of the nonperturbative input in QCD
  resummed Drell-Yan $Q_{T}$ distributions}},  {\em Phys. Rev.} {\bf D63}
  (2001) 114011, [\href{http://arxiv.org/abs/hep-ph/0012348}{{\tt
  hep-ph/0012348}}].

\bibitem{Catani:2015vma}
S.~Catani, D.~de~Florian, G.~Ferrera, and M.~Grazzini, {\it {Vector boson
  production at hadron colliders: transverse-momentum resummation and leptonic
  decay}},  {\em JHEP} {\bf 12} (2015) 047,
  [\href{http://arxiv.org/abs/1507.06937}{{\tt arXiv:1507.06937}}].

\bibitem{Ebert:2016gcn}
M.~A. Ebert and F.~J. Tackmann, {\it {Resummation of Transverse Momentum
  Distributions in Distribution Space}},  {\em JHEP} {\bf 02} (2017) 110,
  [\href{http://arxiv.org/abs/1611.08610}{{\tt arXiv:1611.08610}}].

\bibitem{Monni:2016ktx}
P.~F. Monni, E.~Re, and P.~Torrielli, {\it {Higgs Transverse-Momentum
  Resummation in Direct Space}},  {\em Phys. Rev. Lett.} {\bf 116} (2016),
  no.~24 242001, [\href{http://arxiv.org/abs/1604.02191}{{\tt
  arXiv:1604.02191}}].

\bibitem{Ellis:1985er}
R.~K. Ellis and J.~C. Sexton, {\it {QCD Radiative Corrections to Parton Parton
  Scattering}},  {\em Nucl. Phys.} {\bf B269} (1986) 445--484.

\bibitem{Aversa:1988vb}
F.~Aversa, P.~Chiappetta, M.~Greco, and J.~P. Guillet, {\it {QCD Corrections to
  Parton-Parton Scattering Processes}},  {\em Nucl. Phys.} {\bf B327} (1989)
  105.

\bibitem{Dasgupta:2014yra}
M.~Dasgupta, F.~Dreyer, G.~P. Salam, and G.~Soyez, {\it {Small-radius jets to
  all orders in QCD}},  {\em JHEP} {\bf 04} (2015) 039,
  [\href{http://arxiv.org/abs/1411.5182}{{\tt arXiv:1411.5182}}].

\bibitem{Dai:2016hzf}
L.~Dai, C.~Kim, and A.~K. Leibovich, {\it {Fragmentation of a Jet with Small
  Radius}},  {\em Phys. Rev.} {\bf D94} (2016), no.~11 114023,
  [\href{http://arxiv.org/abs/1606.07411}{{\tt arXiv:1606.07411}}].

\bibitem{Dulat:2015mca}
S.~Dulat, T.-J. Hou, J.~Gao, M.~Guzzi, J.~Huston, P.~Nadolsky, J.~Pumplin,
  C.~Schmidt, D.~Stump, and C.~P. Yuan, {\it {New parton distribution functions
  from a global analysis of quantum chromodynamics}},  {\em Phys. Rev.} {\bf
  D93} (2016), no.~3 033006, [\href{http://arxiv.org/abs/1506.07443}{{\tt
  arXiv:1506.07443}}].

\bibitem{Jager:2002xm}
B.~Jager, A.~Schafer, M.~Stratmann, and W.~Vogelsang, {\it {Next-to-leading
  order QCD corrections to high p(T) pion production in longitudinally
  polarized pp collisions}},  {\em Phys. Rev.} {\bf D67} (2003) 054005,
  [\href{http://arxiv.org/abs/hep-ph/0211007}{{\tt hep-ph/0211007}}].

\bibitem{Airapetian:2012ki}
{\bf HERMES} Collaboration, A.~Airapetian et~al., {\it {Multiplicities of
  charged pions and kaons from semi-inclusive deep-inelastic scattering by the
  proton and the deuteron}},  {\em Phys. Rev.} {\bf D87} (2013) 074029,
  [\href{http://arxiv.org/abs/1212.5407}{{\tt arXiv:1212.5407}}].

\bibitem{Adolph:2013stb}
{\bf COMPASS} Collaboration, C.~Adolph et~al., {\it {Hadron Transverse Momentum
  Distributions in Muon Deep Inelastic Scattering at 160 GeV/$c$}},  {\em Eur.
  Phys. J.} {\bf C73} (2013), no.~8 2531,
  [\href{http://arxiv.org/abs/1305.7317}{{\tt arXiv:1305.7317}}]. [Erratum:
  Eur. Phys. J.C75,no.2,94(2015)].

\bibitem{Boglione:2016bph}
M.~Boglione, J.~Collins, L.~Gamberg, J.~O. Gonzalez-Hernandez, T.~C. Rogers,
  and N.~Sato, {\it {Kinematics of Current Region Fragmentation in
  Semi-Inclusive Deeply Inelastic Scattering}},  {\em Phys. Lett.} {\bf B766}
  (2017) 245--253, [\href{http://arxiv.org/abs/1611.10329}{{\tt
  arXiv:1611.10329}}].

\bibitem{Kang:2015msa}
Z.-B. Kang, A.~Prokudin, P.~Sun, and F.~Yuan, {\it {Extraction of Quark
  Transversity Distribution and Collins Fragmentation Functions with QCD
  Evolution}},  {\em Phys. Rev.} {\bf D93} (2016), no.~1 014009,
  [\href{http://arxiv.org/abs/1505.05589}{{\tt arXiv:1505.05589}}].

\bibitem{Su:2014wpa}
P.~Sun, J.~Isaacson, C.~P. Yuan, and F.~Yuan, {\it {Universal Non-perturbative
  Functions for SIDIS and Drell-Yan Processes}},
  \href{http://arxiv.org/abs/1406.3073}{{\tt arXiv:1406.3073}}.

\bibitem{Bacchetta:2017gcc}
A.~Bacchetta, F.~Delcarro, C.~Pisano, M.~Radici, and A.~Signori, {\it
  {Extraction of partonic transverse momentum distributions from semi-inclusive
  deep-inelastic scattering, Drell-Yan and Z-boson production}},
  \href{http://arxiv.org/abs/1703.10157}{{\tt arXiv:1703.10157}}.

\bibitem{Balazs:1997hv}
C.~Balazs, E.~L. Berger, S.~Mrenna, and C.~P. Yuan, {\it {Photon pair
  production with soft gluon resummation in hadronic interactions}},  {\em
  Phys. Rev.} {\bf D57} (1998) 6934--6947,
  [\href{http://arxiv.org/abs/hep-ph/9712471}{{\tt hep-ph/9712471}}].

\bibitem{Balazs:2000wv}
C.~Balazs and C.~P. Yuan, {\it {Higgs boson production at the LHC with soft
  gluon effects}},  {\em Phys. Lett.} {\bf B478} (2000) 192--198,
  [\href{http://arxiv.org/abs/hep-ph/0001103}{{\tt hep-ph/0001103}}].

\bibitem{Balazs:2007hr}
C.~Balazs, E.~L. Berger, P.~M. Nadolsky, and C.~P. Yuan, {\it {Calculation of
  prompt diphoton production cross-sections at Tevatron and LHC energies}},
  {\em Phys. Rev.} {\bf D76} (2007) 013009,
  [\href{http://arxiv.org/abs/0704.0001}{{\tt arXiv:0704.0001}}].

\bibitem{deFlorian:2007ekg}
D.~de~Florian, R.~Sassot, and M.~Stratmann, {\it {Global analysis of
  fragmentation functions for protons and charged hadrons}},  {\em Phys. Rev.}
  {\bf D76} (2007) 074033, [\href{http://arxiv.org/abs/0707.1506}{{\tt
  arXiv:0707.1506}}].

\bibitem{Vogt:2004ns}
A.~Vogt, {\it {Efficient evolution of unpolarized and polarized parton
  distributions with QCD-PEGASUS}},  {\em Comput. Phys. Commun.} {\bf 170}
  (2005) 65--92, [\href{http://arxiv.org/abs/hep-ph/0408244}{{\tt
  hep-ph/0408244}}].

\bibitem{Anderle:2015lqa}
D.~P. Anderle, F.~Ringer, and M.~Stratmann, {\it {Fragmentation Functions at
  Next-to-Next-to-Leading Order Accuracy}},  {\em Phys. Rev.} {\bf D92} (2015),
  no.~11 114017, [\href{http://arxiv.org/abs/1510.05845}{{\tt
  arXiv:1510.05845}}].

\bibitem{Bodwin:2014gia}
G.~T. Bodwin, H.~S. Chung, U.-R. Kim, and J.~Lee, {\it {Fragmentation
  contributions to $J/\psi$ production at the Tevatron and the LHC}},  {\em
  Phys. Rev. Lett.} {\bf 113} (2014), no.~2 022001,
  [\href{http://arxiv.org/abs/1403.3612}{{\tt arXiv:1403.3612}}].

\bibitem{deFlorian:2014xna}
D.~de~Florian, R.~Sassot, M.~Epele, R.~J. Hernandez-Pinto, and M.~Stratmann,
  {\it {Parton-to-Pion Fragmentation Reloaded}},  {\em Phys. Rev.} {\bf D91}
  (2015), no.~1 014035, [\href{http://arxiv.org/abs/1410.6027}{{\tt
  arXiv:1410.6027}}].

\bibitem{Collins:2016hqq}
J.~Collins, L.~Gamberg, A.~Prokudin, T.~C. Rogers, N.~Sato, and B.~Wang, {\it
  {Relating Transverse Momentum Dependent and Collinear Factorization Theorems
  in a Generalized Formalism}},  {\em Phys. Rev.} {\bf D94} (2016), no.~3
  034014, [\href{http://arxiv.org/abs/1605.00671}{{\tt arXiv:1605.00671}}].

\bibitem{Boglione:2014oea}
M.~Boglione, J.~O. Gonzalez~Hernandez, S.~Melis, and A.~Prokudin, {\it {A study
  on the interplay between perturbative QCD and CSS/TMD formalism in SIDIS
  processes}},  {\em JHEP} {\bf 02} (2015) 095,
  [\href{http://arxiv.org/abs/1412.1383}{{\tt arXiv:1412.1383}}].

\bibitem{Dasgupta:2001sh}
M.~Dasgupta and G.~Salam, {\it {Resummation of nonglobal QCD observables}},
  {\em Phys.Lett.} {\bf B512} (2001) 323--330,
  [\href{http://arxiv.org/abs/hep-ph/0104277}{{\tt hep-ph/0104277}}].

\bibitem{Banfi:2002hw}
A.~Banfi, G.~Marchesini, and G.~Smye, {\it {Away from jet energy flow}},  {\em
  JHEP} {\bf 0208} (2002) 006, [\href{http://arxiv.org/abs/hep-ph/0206076}{{\tt
  hep-ph/0206076}}].

\bibitem{Dasgupta:2012hg}
M.~Dasgupta, K.~Khelifa-Kerfa, S.~Marzani, and M.~Spannowsky, {\it {On jet mass
  distributions in Z+jet and dijet processes at the LHC}},  {\em JHEP} {\bf
  1210} (2012) 126, [\href{http://arxiv.org/abs/1207.1640}{{\tt
  arXiv:1207.1640}}].

\bibitem{Larkoski:2015zka}
A.~J. Larkoski, I.~Moult, and D.~Neill, {\it {Non-Global Logarithms,
  Factorization, and the Soft Substructure of Jets}},  {\em JHEP} {\bf 09}
  (2015) 143, [\href{http://arxiv.org/abs/1501.04596}{{\tt arXiv:1501.04596}}].

\bibitem{Larkoski:2016zzc}
A.~J. Larkoski, I.~Moult, and D.~Neill, {\it {The Analytic Structure of
  Non-Global Logarithms: Convergence of the Dressed Gluon Expansion}},  {\em
  JHEP} {\bf 11} (2016) 089, [\href{http://arxiv.org/abs/1609.04011}{{\tt
  arXiv:1609.04011}}].

\bibitem{Neill:2016stq}
D.~Neill, {\it {The Asymptotic Form of Non-Global Logarithms, Black Disc
  Saturation, and Gluonic Deserts}},  {\em JHEP} {\bf 01} (2017) 109,
  [\href{http://arxiv.org/abs/1610.02031}{{\tt arXiv:1610.02031}}].

\bibitem{Becher:2016mmh}
T.~Becher, M.~Neubert, L.~Rothen, and D.~Y. Shao, {\it {Factorization and
  Resummation for Jet Processes}},  {\em JHEP} {\bf 11} (2016) 019,
  [\href{http://arxiv.org/abs/1605.02737}{{\tt arXiv:1605.02737}}].

\bibitem{Caron-Huot:2015bja}
S.~Caron-Huot, {\it {Resummation of non-global logarithms and the BFKL
  equation}},  \href{http://arxiv.org/abs/1501.03754}{{\tt arXiv:1501.03754}}.

\bibitem{Hagiwara:2015bia}
Y.~Hagiwara, Y.~Hatta, and T.~Ueda, {\it {Hemisphere jet mass distribution at
  finite $N_c$}},  {\em Phys. Lett.} {\bf B756} (2016) 254--258,
  [\href{http://arxiv.org/abs/1507.07641}{{\tt arXiv:1507.07641}}].

\end{thebibliography}\endgroup
\end{document}